%% file: CCInclNukeTarget.tex
\newcommand{\DoPrePrint}{0} % 0 for 2-column submission/review format; 1 for double-spaced, line-numbered preprint
\newcommand{\sizecheck}{0} % 0 to do nothing; 1 to check size
\newcommand{\PRLsupp}{0}   % 0 to add addendix to main file; 1 to put appendix in separate supplemental material document
\newcommand{\DoNotes}{0} % 0 to hide author's notes (\note{}).  1 to show notes.  notes should be removed entirely from text before submission.
\newcommand{\DoColor}{0} % 0 to do BW, 1 to do color
\newcommand{\version}{38v7} %define the version to show in preprint
\ifnum\DoPrePrint=1
  \RequirePackage{lineno}  % LINENO  CTOBS
  \documentclass[aps,preprint,showpacs,superscriptaddress,linenumbers]{revtex4} % CTOBS
  \usepackage{lineno}
\else
  \documentclass[aps,prl,twocolumn,showpacs,superscriptaddress,nofootinbib]{revtex4}  % UTOBS
\fi

\usepackage{amsmath}    % need for subequations
\usepackage{graphicx}   % for figures
\usepackage{xspace}
\usepackage{units}
\usepackage{hyperref}

\newcommand{\minerva}{MINERvA\xspace}  %use a v instead of nu
\newcommand{\minos}{MINOS\xspace}

\newcommand{\numi}{NuMI\xspace}
\newcommand{\numu}{\ensuremath{\nu_{\mu}}\xspace}
\newcommand{\numubar}{\ensuremath{\bar{\nu}_{\mu}}\xspace}

\newcommand{\GeV}{\ensuremath{\mbox{GeV}}\xspace}

\newcommand{\GeVcsq}{\ensuremath{\mbox{GeV}^2}\xspace}
\newcommand{\cmsq}{\ensuremath{\mbox{cm}^2}\xspace}
\newcommand{\Xbj}{\ensuremath{x}\xspace}
\newcommand{\Enu}{\ensuremath{E_{\nu}}\xspace}

\newcommand{\A}{\ensuremath{A}\xspace}  %nuclear target number
\newcommand{\Emu}{\ensuremath{E_{\mu}}\xspace}
\newcommand{\thetamu}{\ensuremath{\theta_{\mu}}\xspace}
\newcommand{\mum}{\ensuremath{\mu^{-}}\xspace}

\newcommand{\recoilE}{\ensuremath{\nu}\xspace}
\newcommand{\Qsq}{\ensuremath{Q^{2}}\xspace}

\ifnum\PRLsupp=0  %for arxiv
\newcommand{\SuppLocation}{ }
\newcommand{\SuppNote}[1]{\footnote{#1.}}
\else             %for PRL
\newcommand{\SuppLocation}{~\cite{SupplementalMaterial}}
\newcommand{\SuppNote}[1]{ (#1)}
\fi

\ifnum\DoColor=1
\graphicspath{ {./figures-color/} }
\else
\graphicspath{ {./figures/} }
\fi

\newif\ifpdf
\ifx\pdfoutput\undefined
\pdffalse
\else
\pdfoutput=1
\pdftrue
\fi
\ifpdf
\usepackage{graphicx}
\usepackage{epstopdf}
\else
\usepackage{graphicx}
\fi

\ifnum\DoNotes=1
\providecommand{\note}[1]
{\marginpar{\LARGE $\spadesuit$}$\spadesuit$ {\bf #1} $\spadesuit$}
\else
\providecommand{\note}[1]
{}
\fi % notes

\begin{document}

\ifnum\DoPrePrint=1
\linenumbers %              CTOBS

\begin{flushright}
  {\large \minerva Publication \version} \\
\end{flushright}
\fi

\title{Measurement of Ratios of $\nu_{\mu}$ Charged-Current Cross Sections \\ on C, Fe, and Pb to CH at Neutrino Energies 2--20~GeV}

%%% This is the neutrino communications author list
%%%
%%% This file created automagically via Glaucus using this URL:
%%% https://neutrino.otterbein.edu/Glaucus/web/latex_paper.cgi?paper_id=30
%%%
%% MANUAL PARTS OF AUTHOR LIST

%% (1) need to add ``\thanks{\deceased}'' after DeMaat, Gobbi, Tzanakos
\newcommand{\deceased}{Deceased}

%% (2) we offered Jan authorship, so here it is
%% have to put his author line in by hand
\newcommand{\wroclaw}{Institute of Theoretical Physics, Wroc\l aw University, Wroc\l aw, Poland}   
%% List of institution addresses, in command form.
%% List of institution addresses, in command form.
\newcommand{\Rutgers}{Rutgers, The State University of New Jersey, Piscataway, New Jersey 08854, USA}
\newcommand{\Hampton}{Hampton University, Dept. of Physics, Hampton, VA 23668, USA}
\newcommand{\Dortmund}{Institute of Physics, Dortmund University, 44221, Germany }
\newcommand{\Otterbein}{Department of Physics, Otterbein University, 1 South Grove Street, Westerville, OH, 43081 USA}
\newcommand{\JMU}{James Madison University, Harrisonburg, Virginia 22807, USA}
\newcommand{\Florida}{University of Florida, Department of Physics, Gainesville, FL 32611}
\newcommand{\UCIrvine}{Department of Physics and Astronomy, University of California, Irvine, Irvine, California 92697-4575, USA}
\newcommand{\CBPF}{Centro Brasileiro de Pesquisas F\'{i}sicas, Rua Dr. Xavier Sigaud 150, Urca, Rio de Janeiro, RJ, 22290-180, Brazil}
\newcommand{\PUCP}{Secci\'{o}n F\'{i}sica, Departamento de Ciencias, Pontificia Universidad Cat\'{o}lica del Per\'{u}, Apartado 1761, Lima, Per\'{u}}
\newcommand{\INRM}{Institute for Nuclear Research of the Russian Academy of Sciences, 117312 Moscow, Russia}
\newcommand{\Jlab}{Jefferson Lab, 12000 Jefferson Avenue, Newport News, VA 23606, USA}
\newcommand{\Pittsburgh}{Department of Physics and Astronomy, University of Pittsburgh, Pittsburgh, Pennsylvania 15260, USA}
\newcommand{\Guanajuato}{Campus Le\'{o}n y Campus Guanajuato, Universidad de Guanajuato, Lascurain de Retana No. 5, Col. Centro. Guanajuato 36000, Guanajuato M\'{e}xico.}
\newcommand{\Athens}{Department of Physics, University of Athens, GR-15771 Athens, Greece}
\newcommand{\Tufts}{Physics Department, Tufts University, Medford, Massachusetts 02155, USA}
\newcommand{\WM}{Department of Physics, College of William \& Mary, Williamsburg, Virginia 23187, USA}
\newcommand{\FNAL}{Fermi National Accelerator Laboratory, Batavia, Illinois 60510, USA}
\newcommand{\Purdue}{Department of Chemistry and Physics, Purdue University Calumet, Hammond, Indiana 46323, USA}
\newcommand{\MCLA}{Massachusetts College of Liberal Arts, 375 Church Street, North Adams, MA 01247}
\newcommand{\UMD}{Department of Physics, University of Minnesota -- Duluth, Duluth, Minnesota 55812, USA}
\newcommand{\Northwestern}{Northwestern University, Evanston, Illinois 60208}
\newcommand{\UNI}{Universidad Nacional de Ingenier\'{i}a, Apartado 31139, Lima, Per\'{u}}
\newcommand{\Rochester}{University of Rochester, Rochester, New York 14610 USA}
\newcommand{\Austin}{Department of Physics, University of Texas, 1 University Station, Austin, Texas 78712, USA}
\newcommand{\USM}{Departamento de F\'{i}sica, Universidad T\'{e}cnica Federico Santa Mar\'{i}a, Avda. Espa\~{n}a 1680 Casilla 110-V, Valpara\'{i}so, Chile}
\newcommand{\Geneva}{University of Geneva, Geneva, Switzerland}
\newcommand{\Chicago}{Enrico Fermi Institute, University of Chicago, Chicago, IL 60637 USA}
\newcommand{\hired}{}
\newcommand{\giulianoThanks}{now at Vrije Universiteit Brussel, Pleinlaan 2, B-1050 Brussels, Belgium}
\newcommand{\LazaThanks}{also at Department of Physics, University of Antananarivo, Madagascar}
\newcommand{\ticeThanks}{now at Argonne National Laboratory, Argonne, IL 60439, USA }

% 78 total signatories.
%%%%%%%%%
% User edit area for lead authors
% Tice, Datta, Mouseau lead authors
%tice affiliation includes ANL for this paper only must not use tice thanks
\author{B.G.~Tice}\thanks{\ticeThanks}    \affiliation{\Rutgers}
\author{M.~Datta}                         \affiliation{\Hampton}
\author{J.~Mousseau}                      \affiliation{\Florida}
% back to automatic list below
\author{L.~Aliaga}                        \affiliation{\WM}  \affiliation{\PUCP}
\author{O.~Altinok}                       \affiliation{\Tufts}
\author{M.G.~Barrios~Sazo}                \affiliation{\Guanajuato}
\author{M.~Betancourt}                    \affiliation{\FNAL}
\author{A.~Bodek}                         \affiliation{\Rochester}
\author{A.~Bravar}                        \affiliation{\Geneva}
\author{W.K.~Brooks}                      \affiliation{\USM}
\author{H.~Budd}                          \affiliation{\Rochester}
\author{M.~J.~Bustamante~}                \affiliation{\PUCP}
\author{A.~Butkevich}                     \affiliation{\INRM}
\author{D.A.~Martinez~Caicedo}            \affiliation{\CBPF}  \affiliation{\FNAL}
\author{C.M.~Castromonte}                 \affiliation{\CBPF}
\author{M.E.~Christy}                     \affiliation{\Hampton}
\author{J.~Chvojka}                       \affiliation{\Rochester}
\author{H.~da~Motta}                      \affiliation{\CBPF}
\author{J.~Devan}                         \affiliation{\WM}
\author{S.A.~Dytman}                      \affiliation{\Pittsburgh}
\author{G.A.~D\'{i}az~}                   \affiliation{\PUCP}
\author{B.~Eberly}                        \affiliation{\Pittsburgh}
\author{J.~Felix}                         \affiliation{\Guanajuato}
\author{L.~Fields}                        \affiliation{\Northwestern}
\author{G.A.~Fiorentini}                  \affiliation{\CBPF}
\author{A.M.~Gago}                        \affiliation{\PUCP}
\author{H.~Gallagher}                     \affiliation{\Tufts}
\author{R.~Gran}                          \affiliation{\UMD}
\author{D.A.~Harris}                      \affiliation{\FNAL}
\author{A.~Higuera}                       \affiliation{\Guanajuato}
\author{K.~Hurtado}                       \affiliation{\CBPF}  \affiliation{\UNI}
\author{M.~Jerkins}                       \affiliation{\Austin}
\author{T.~Kafka}                         \affiliation{\Tufts}
\author{M.~Kordosky}                      \affiliation{\WM}
\author{S.A.~Kulagin}                     \affiliation{\INRM}
\author{T.~Le}                            \affiliation{\Rutgers}
\author{G.~Maggi}\thanks{\giulianoThanks}  \affiliation{\USM}
\author{E.~Maher}                         \affiliation{\MCLA}
\author{S.~Manly}                         \affiliation{\Rochester}
\author{W.A.~Mann}                        \affiliation{\Tufts}
\author{C.M.~Marshall}                    \affiliation{\Rochester}
\author{C.~Martin~Mari}                   \affiliation{\Geneva}
\author{K.S.~McFarland}                   \affiliation{\Rochester}  \affiliation{\FNAL}
\author{C.L.~McGivern}                    \affiliation{\Pittsburgh}
\author{A.M.~McGowan}                     \affiliation{\Rochester}
\author{J.~Miller}                        \affiliation{\USM}
\author{A.~Mislivec}                      \affiliation{\Rochester}
\author{J.G.~Morf\'{i}n}                  \affiliation{\FNAL}
\author{T.~Muhlbeier}                     \affiliation{\CBPF}
\author{D.~Naples}                        \affiliation{\Pittsburgh}
\author{J.K.~Nelson}                      \affiliation{\WM}
\author{A.~Norrick}                       \affiliation{\WM}
\author{J.~Osta}                          \affiliation{\FNAL}
\author{J.L.~Palomino}                    \affiliation{\CBPF}
\author{V.~Paolone}                       \affiliation{\Pittsburgh}
\author{J.~Park}                          \affiliation{\Rochester}
\author{C.E.~Patrick}                     \affiliation{\Northwestern}
\author{G.N.~Perdue}                      \affiliation{\FNAL}  \affiliation{\Rochester}
\author{L.~Rakotondravohitra}\thanks{\LazaThanks}  \affiliation{\FNAL}
\author{R.D.~Ransome}                     \affiliation{\Rutgers}
\author{H.~Ray}                           \affiliation{\Florida}
\author{L.~Ren}                           \affiliation{\Pittsburgh}
\author{P.A.~Rodrigues}                   \affiliation{\Rochester}
\author{D.~G.~Savage}                     \affiliation{\FNAL}
\author{H.~Schellman}                     \affiliation{\Northwestern}
\author{D.W.~Schmitz}                     \affiliation{\Chicago}
\author{C.~Simon}                         \affiliation{\UCIrvine}
\author{F.D.~Snider}                      \affiliation{\FNAL}
\author{C.J.~Solano~Salinas}              \affiliation{\UNI}
\author{N.~Tagg}                          \affiliation{\Otterbein}
\author{E.~Valencia}                      \affiliation{\Guanajuato}
\author{J.P.~Vel\'{a}squez}               \affiliation{\PUCP}
\author{T.~Walton}                        \affiliation{\Hampton}
\author{J.~Wolcott}                       \affiliation{\Rochester}
\author{G.~Zavala}                        \affiliation{\Guanajuato}
\author{D.~Zhang}                         \affiliation{\WM}
\author{B.P.~Ziemer}                       \affiliation{\UCIrvine}

%%%%%% AUTOMATIC LIST (EDITED AS ABOVE)
%%%%\author{L.Y.~Zhu}                         \affiliation{\Hampton}
%%%%\author{B.P.~Ziemer}                      \affiliation{\UCIrvine}
%% END AUTOMATIC PART
\collaboration{\minerva Collaboration}\ \noaffiliation
\date{\today}
\begin{abstract}
We present measurements of \numu charged-current cross section ratios on carbon, iron, and lead relative to a scintillator (CH) using the fine-grained \minerva detector exposed to the \numi neutrino beam at Fermilab.
The measurements utilize events of energies $2<\Enu<20$~\GeV, with $\left<\Enu\right>=8~\GeV$, which have a reconstructed \mum scattering angle less than $17^\circ$ to extract ratios of inclusive total cross sections as a function of neutrino energy \Enu and flux-integrated differential cross sections with respect to the Bjorken scaling variable \Xbj.
These results provide the first high-statistics direct measurements of nuclear effects in neutrino scattering using different targets in the same neutrino beam.
Measured cross section ratios exhibit a relative depletion at low \Xbj and enhancement at large \Xbj.
Both become more pronounced as the nucleon number of the target nucleus increases.
The data are not reproduced by GENIE, a conventional neutrino-nucleus scattering simulation, or by the alternative models for the nuclear dependence of inelastic scattering that are considered.
\end{abstract}

\ifnum\sizecheck=0
\maketitle
\fi

\input{intro-NukeCC.tex}

\input{data-and-analysis-NukeCC.tex}

\input{selection-NukeCC.tex}
\input{CrossSectionRatio-NukeCC.tex}

\input{systematic-NukeCC.tex}
\input{results-NukeCC.tex}

\ifnum\sizecheck=0
\input{acknowledgments.tex}
\input{biblio-NukeCC-bibtex.tex}
\fi

%don't add appendix to main if doing a separate supplement
\ifnum\PRLsupp=0
\clearpage
\onecolumngrid
\input{appendix-NukeCC.tex}

\fi

\end{document}

%% file: intro-NukeCC.tex
Measurements of charged lepton scattering
from different nuclei show that the cross section ratio on a heavy nucleus relative to 
the deuteron $\sigma^{\A}/\sigma^{D}$ deviates from unity by as much as 20\%. 
This demonstrates nontrivial nuclear effects over a wide range of Bjorken's scaling variable \Xbj~\cite{Arneodo:1992wf,geesaman1995nuclear,Norton:2003cb,rith2014}.
These observations, first reported by the European Muon Collaboration (EMC)~\cite{Aubert:1983xm,Aubert1987740} in 1983, signal a difference in the quark-parton structure of bound nucleons from that of free nucleons
and have triggered theoretical exploration of
background nuclear mechanisms~\cite{Arneodo:1992wf,Norton:2003cb}.

In neutrino physics, understanding nuclear effects is necessary for correct
interpretation of measurements of electroweak parameters 
and evaluation of corresponding uncertainties~\cite{PhysRevLett.88.091802}.
The precision of modern neutrino oscillation experiments
has rekindled interest in measuring nuclear effects, albeit at lower 
neutrino energies where elastic and resonance processes, rather than deep inelastic processes, are dominant~\cite{Coloma:2013tba}.

Neutrino scattering, unlike that of charged leptons, involves the axial-vector current and
is sensitive to specific quark and antiquark flavors.
Therefore, nuclear modifications of neutrino cross sections may differ
from those of charged leptons~\cite{Kulagin2006126,PhysRevC.84.054610,Haider:2013eqa,Kopeliovich:2012kw}.
An indirect extraction of neutrino deep inelastic structure function ratios using NuTeV Fe~\cite{PhysRevD.74.012008} and
CHORUS Pb~\cite{onegut200665} data suggests this is the case~\cite{Kovarik:2010uv}.
If confirmed, this either challenges the validity of QCD factorization for processes
involving bound nucleons or signals inconsistency between neutrino and charged lepton data.
Another study~\cite{PhysRevLett.110.212301} using different techniques does not find this behavior.
Neutrino scattering data are necessary for separation of valence and sea quark contributions to parton distribution functions~\cite{PhysRevC.76.065207,PhysRevD.77.054013,Kovarik:2010uv}, but high-statistics data from iron and lead must be corrected to account for poorly measured nuclear modifications.

Direct measurements of neutrino cross section ratios for different nuclei
are therefore of significant interest and importance.
So far, the only such measurements are ratios of Ne to D~\cite{Parker19841,Cooper1984133,PhysRevD.32.2441}, 
but these are rarely used because of large statistical uncertainties and model-dependent extraction from a mixed H-Ne target.
In this Letter, we report the first measurement of inclusive charged-current neutrino cross section
ratios of C, Fe, and Pb to scintillator (CH) as functions of neutrino energy \Enu and \Xbj.
This is the first application to neutrino physics of the EMC-style technique of measuring nuclear dependence with multiple nuclear targets in the same beam and detector.

%% file: data-and-analysis-NukeCC.tex
\minerva uses a finely segmented detector to record interactions of neutrinos produced by the NuMI beam line~\cite{Anderson:1998zza} at Fermilab.  
Data for this analysis come from $2.94\times 10^{20}$ protons on target taken between March 2010 and April 2012 
when the beam line produced a broadband neutrino beam peaked at \unit[3.5]{\GeV} with $>95\%$ \numu at the peak energy.
The \minerva detector is comprised of 120 hexagonal modules 
perpendicular to the $z$ axis, which is tilted \unit[58]{mrad} upwards with respect to the beam line~\cite{minerva_nim}.
There are four module types: active tracking, 
electromagnetic calorimeter, hadronic calorimeter, and inactive nuclear target.
The most upstream part of
the detector includes five inactive targets, numbered from upstream to downstream, 
each separated by four active tracking modules.
Target 4 is lead; other targets comprise two or three materials arranged at differing transverse positions filling the $x-y$ plane.
Targets 1, 2, and 5 are constructed of steel and lead plates joined together;
target 3 has graphite, steel, and lead plates.  
Total fiducial masses of C, Fe, and Pb in the nuclear target region 
are 0.159, 0.628, and 0.711 tons, respectively.
A fully active tracking region with a fiducial mass of 5.48 tons is downstream of the nuclear target region.
The target and tracker regions are surrounded by electromagnetic and hadronic calorimeters.
The MINOS near detector, a magnetized iron spectrometer~\cite{Michael:2008bc}, is located 
\unit[2]{m} downstream of the \minerva detector.

Neutrino flux is predicted using a GEANT4-based simulation tuned to hadron production data~\cite{Alt:2006fr} as described in Ref.~\cite{nubarprl}\SuppNote{See Supplemental Material\SuppLocation for a table of the simulated neutrino flux}.
Neutrino interactions in the detector are simulated using GENIE 2.6.2~\cite{Andreopoulos201087}. 
In GENIE, the initial nucleon momentum is selected from distributions in Refs.~\cite{Bodek:1980ar,Bodek:1981wr}.   
Scattering kinematics are calculated in the (off-shell) nucleon rest frame.    
The quasielastic cross section is reduced to account for Pauli blocking.    
For quasielastic and resonance processes, free nucleon form factors are used.  
Quasielastic model details are given in Ref.~\cite{nubarprl}. 
Kinematics for nonresonant inelastic processes are selected from the model of Ref.~\cite{Bodek:2004pc}
which effectively includes target mass and higher twist corrections. 
An empirical correction factor based on charged lepton deep inelastic scattering measurements of 
$F_{2}^{D}/F_2^{\left(n+p\right)}$ and $F_{2}^{Fe}/F_{2}^{D}$ is applied to all structure functions 
as a function of \Xbj, independent of the four-momentum transfer squared \Qsq and \A.  
This accounts for all nuclear effects except those related to neutron excess, which are applied separately.

The \minerva detector's response is simulated by a tuned 
GEANT4-based~\cite{Agostinelli2003250,1610988} simulation.  
The energy scale of the detector is set by ensuring
both detected photon statistics and reconstructed energy deposited by
momentum-analyzed throughgoing muons agree in data and simulation.
Calorimetric constants applied to reconstruct the recoil energy 
are determined by simulation. 
This procedure is cross-checked by comparing data and simulation 
of a scaled-down version of the \minerva detector in a 
low energy hadron test beam~\cite{minerva_nim}.

%% file: selection-NukeCC.tex
Charged-current \numu events must have a reconstructed \mum.
The muon is identified by a minimum ionizing track that
traverses \minerva~\cite{minerva_nim} and travels downstream into the
\minos spectrometer~\cite{Michael:2008bc} where its momentum
and charge are measured.
Muon selection and energy (\Emu) reconstruction are described in Refs.~\cite{nubarprl,nuprl,minerva_nim}.
Requiring a matching track in \minos restricts muon acceptance.
To minimize acceptance differences across the \minerva detector, the analysis
requires neutrino energies above \unit[2]{GeV} and muon angles with respect to the beam (\thetamu) 
less than \unit[17]{$^{\circ}$}.
A \unit[20]{GeV} upper limit on neutrino energy reduces the \numubar background to below 1\%.
After all selection criteria, 5953 events in C, $19\,024$ in Fe, $23\,967$ in Pb, and $189\,168$ in CH are analyzed.

The event vertex is the location of the most upstream energy deposition on the muon track
when only one track is reconstructed; a Kalman filter~\cite{Fruhwirth1987444,Luchsinger1993263} is used to fit the vertex position for events with multiple tracks.
Between 10\% and 20\% of selected events in the different samples have a well-reconstructed multitrack vertex; the remainder are single track
or have a poorly reconstructed vertex position based on the $\chi^2$ of the vertex fit.
Events with vertices in targets 2 through 5 and the fully active tracking volume are considered. 
The target 1 sample has a higher background from interactions upstream of the detector.

Events with a vertex in the active tracking region are divided into three statistically independent CH samples used to form ratios with C, Fe, and Pb.
Events are associated with the C, Fe, or Pb of a nuclear target if the vertex position is between one plane upstream and two planes downstream of that nuclear target module and more than 
 \unit[2.5]{cm} away transversely from seams that join different materials in the target.
In single-track events, the muon track is propagated to the longitudinal center of the nuclear target to estimate the vertex 
position and momentum of the muon.
After all cuts, charged-current event selection efficiency ranges from 24\% in the most upstream 
targets to 50\% in the most downstream. 
The large efficiency variation exists because the upstream region has more inert material and smaller \minos solid angle coverage.   

Energy of the hadronic recoil system \recoilE is determined from the calorimetric sum
of energy deposits not associated with the muon track. 
We consider deposits which occur
between \unit[20]{ns} before and \unit[35]{ns} after the muon to reduce contributions from overlap with other neutrino interactions.
Visible energies are weighted to account for the active fraction of scintillator in different regions of the detector.
The overall calorimetric scale comes from fitting reconstructed \recoilE 
to generated \recoilE for simulated events in the active tracking region~\cite{minerva_nim}.
Using the same procedure, additional calorimetric scales for events in 
targets 2 through 5 are obtained as relative to the tracker;
these are, respectively, 1.11, 1.04, 0.99, and 0.98.  

Kinematic variables \Enu, \Xbj, and \Qsq are obtained 
from reconstructed \Emu, \thetamu, and \recoilE:
$\Enu = \Emu + \recoilE$, $\Qsq = 4 \Enu \Emu {\rm{sin}}^2(\frac{\thetamu}{2})$, 
and $\Xbj = \frac{\Qsq}{2 M_{N}\recoilE}$, where $M_{N}$ is the average 
of proton and neutron masses.
Reconstructed \Enu distributions are corrected for detector smearing 
using iterative Bayesian
unfolding~\cite{D'Agostini:1994zf} with four iterations to produce 
event yields as functions of unfolded \Enu, with generated \Enu values from GENIE.

Reconstructed \Xbj is smeared broadly, especially at high \Xbj where quasielastic processes dominate.
For these events, the hadronic recoil system can be a single nucleon, which is not reconstructed well under a calorimetric assumption.
Such significant smearing would cause large uncertainties in the unfolding procedure.
We therefore report cross section ratios as functions of reconstructed \Xbj\SuppNote{See Supplemental Material\SuppLocation for migration matrices necessary for comparisons of theoretical models to these data}.

\begin{figure}[tp]
\centering
\includegraphics[width=0.825\columnwidth]{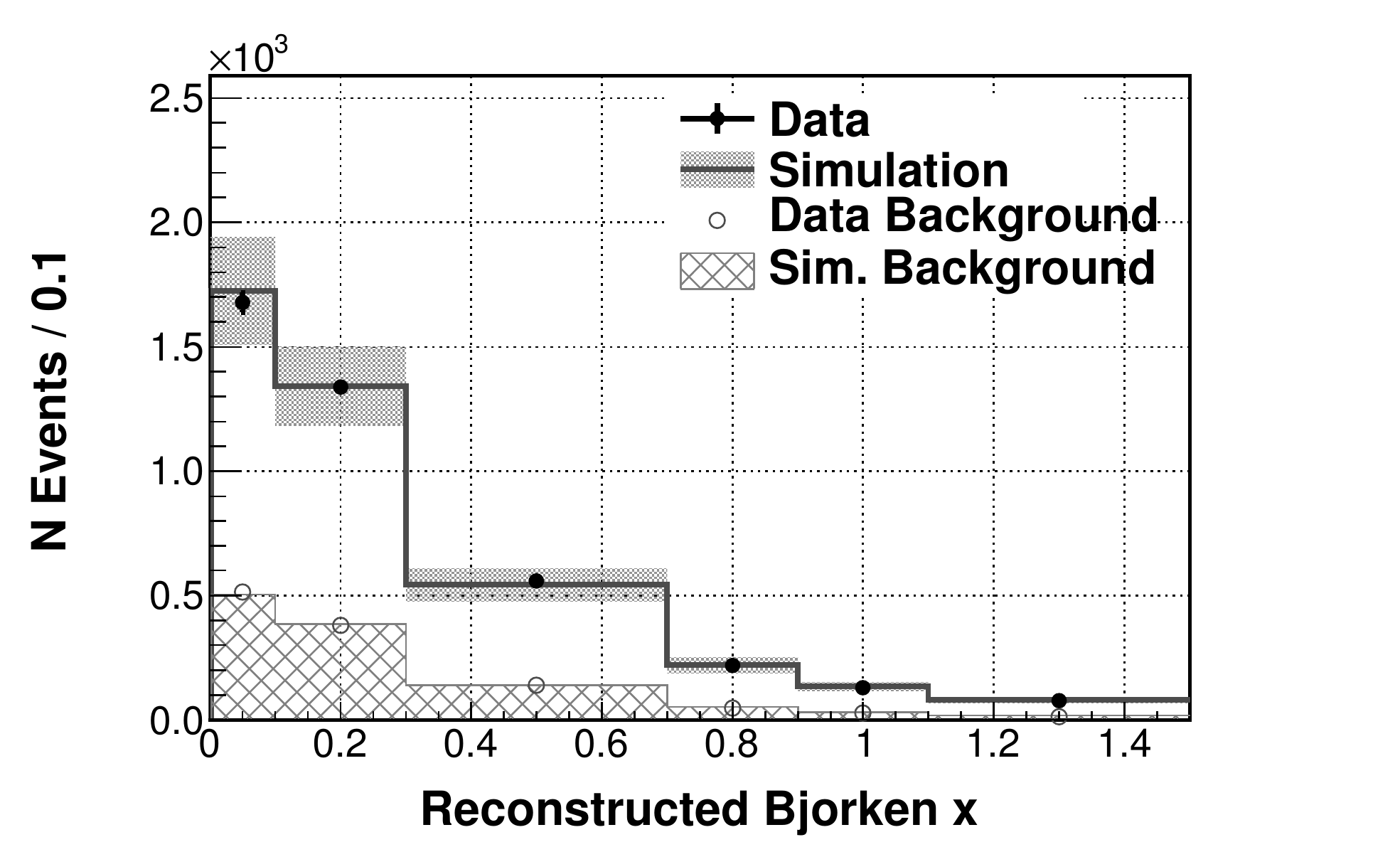}
\caption{Reconstructed Bjorken $x$ distributions in data and simulation for selected inclusive \numu events in the lead of Target 2. The plot includes CH contamination separately estimated using data and simulated events in the tracker region.  Both simulation distributions are normalized to the data by the number of events passing all event selection criteria.  Events are scaled to a bin size of 0.1.  Events with \Xbj greater than 1.5 are not shown.} 
\label{fig:sample_x}
\end{figure}

Nuclear target samples contain events from adjacent tracking modules due to the loose cut on longitudinal vertex position.
This background, called ``CH contamination,''
ranges from 20\% to 40\% and is roughly proportional to 
the ratio of areal densities of the target to surrounding scintillator.
CH contamination
is estimated by extrapolating event rates measured
in the active tracking region to the nuclear target region.
The tracking and nuclear target regions occupy different areas, and therefore have different acceptance into the \minos detector.
Further, the Fe and Pb targets in the nuclear target region stimulates greater activity in hadronic showers, 
which affects tracking efficiency.
To account for the geometric acceptance difference, we apply a 
correction $w^{t,\A}(\Emu,\thetamu)$,
obtained from a large, single-particle simulated \mum sample.
Here $t=2,3,4,5$ is the target number and \A$=\:$C, Fe, Pb is the target nucleus.
We account for $\recoilE$-dependent efficiency differences using
simulated neutrino events to derive a correction $w^{t,\A}(\recoilE)$.
Differences are largest at low \recoilE.
Acceptance- and efficiency-corrected distributions are scaled such that the integrated number of
events in true and estimated backgrounds are equal according to neutrino event simulation.
Figure~\ref{fig:sample_x} shows the \Xbj distribution of events passing all 
selection criteria in data and simulation; the estimated CH contamination is also shown.

Deviations found in simulated events between the estimated CH contamination by extrapolation and the predicted CH contamination using generator-level information
are not fully covered by statistical uncertainty at the 68\% confidence level in all targets.
Additional systematic uncertainty is applied to ensure coverage at the 68\% level.
CH contamination uncertainties are 1\%--8\% from these systematic deviations and 2\%--5\% from statistics.

Small backgrounds from \numubar ($<0.4\%$) and neutral current ($<0.1\%$) events are estimated using simulation and subtracted.
Transverse smearing within a nuclear target causes roughly 0.5\% of the interactions to be assigned an incorrect target nucleus;
this is also estimated by simulation and subtracted.
A background from upstream neutrino interactions of $6.2\%\pm2.4\%$ exists only in target 2 for one third of the beam exposure, 
because two of the modules upstream of target 1 were not yet instrumented;
affected data are weighted accordingly.

GENIE predicts a sample not dominated by any single process.
Table~\ref{tab:sample_comp_Summary_table} shows the predicted prevalence of processes in bins of reconstructed \Xbj.
%%%
\begingroup
\squeezetable
\begin{table}
\begin{tabular}{cccccccc}
\hline \hline
Reconstructed \Xbj & I    & II   & III  & IV   & V    & Mean Generated \Qsq\\
                   & (\%) & (\%) & (\%) & (\%) & (\%) &  (\GeVcsq)\\
\hline
0.0--0.1 & 11.3    &  42.5    &  5.9   &  19.2  & 15.7  &  0.23 \\
0.1--0.3 & 13.6    &  36.4    &  16.7  &  9.1   & 23.0  &  0.70 \\
0.3--0.7 & 32.7    &  32.8    &  11.8  &  1.4   & 21.1  &  1.00 \\
0.7--0.9 & 55.1    &  25.4    &  4.3   &  0.5   & 14.6  &  0.95 \\
0.9--1.1 & 62.7    &  21.6    &  2.8   &  0.5   & 12.3  &  0.90 \\
1.1--1.5 & 69.6    &  18.1    &  1.9   &  0.4   & 9.9   &  0.82 \\
$>1.5$   & 79.1    &  12.8    &  0.6   &  0.3   & 7.1   &  0.86 \\ 
\hline \hline
\end{tabular}
\caption{
  Average sample composition of selected nuclear target and tracker events in reconstructed \Xbj bins
  based on GENIE simulation of different physics processes, together with the average generated \Qsq.
  Processes are (I) quasielastic, (II) baryon resonance production, (III) deep inelastic scattering at $\Qsq>1$~\GeVcsq and $W>2$~\GeV, (IV) deep inelastic scattering at $\Qsq<1$~\GeVcsq and $W>2$~\GeV, and (V) nonresonant inelastic continuum with $W<2$~\GeV.
} 
\label{tab:sample_comp_Summary_table}
\end{table}
\endgroup
We compare GENIE's prediction for inclusive cross section ratios restricted to $2<\Enu<20~\GeV$ and $\thetamu<17^{\circ}$ to two other models for nuclear modification of structure functions\SuppNote{See Supplemental Material\SuppLocation for a table summarizing the comparison of models of nuclear modification ofinelastic structure functions}.
The Kulagin-Petti microphysical model starts with neutrino-nucleon structure functions and incorporates \A-dependent nuclear effects~\cite{PhysRevD.76.094023,Kulagin2006126}.
The updated Bodek-Yang treatment~\cite{by13} of the model implemented in GENIE~\cite{Bodek:2004pc} includes an A-dependent empirical correction based on charged lepton measurements on the nuclei of interest.
Although nuclear structure functions vary by 20\% among models, ratios of structure functions in Fe or Pb to C differ by $\lesssim$1\%.

%% file: CrossSectionRatio-NukeCC.tex
%%\subsection{Differential Cross Sections Ratios}

The total cross section for an \Enu bin $i$ is 
$\sigma_{i} = \frac{\Sigma_j U_{ij}(N_j-N_j^{bg})}{\varepsilon_i T \Phi_{i}} $,
where $U_{ij}$ is a matrix that accounts for smearing from true energy bin $i$ to reconstructed
energy bin $j$; $N_{j}$ and $N_{j}^{bg}$ are the numbers of total
and estimated background events in bin $j$, respectively;
$\varepsilon_{i}$ is the efficiency for reconstructing signal events in bin $i$;
$T$ is the number of target nucleons; and $\Phi_{i}$ is the neutrino flux bin $i$.
The flux-integrated differential cross section for a reconstructed $x$ bin $j$ is 
$\left(\frac{d\sigma}{d\Xbj}\right)_{j} = \frac{N_{j}-N_{j}^{bg}}{\varepsilon_{j} T \Phi \Delta_{j}(\Xbj)}$, 
where $\Phi$ is the neutrino flux integrated from $2$ to $20$~\GeV, $\Delta_{j}(\Xbj)$ is bin width, 
and other terms have the same meaning as above.
No correction is applied to account for neutron excess in any target nuclei.

%% file: systematic-NukeCC.tex
%%\section{Systematic estimates}

The main sources of systematic uncertainty in the 
cross section ratio measurements are 
(I) subtraction of CH contamination; 
(II) detector response to muons and hadrons;
(III) neutrino interaction models;
(IV) final state interaction models; 
and (V) target number.
Uncertainty in flux is considered but negligible.
All uncertainties are evaluated by repeating the cross section
analysis with systematic shifts applied to simulation.
%%%
\begingroup
\squeezetable
\begin{table}
  \begin{tabular}{cccccccc}
    \hline \hline
    \Xbj & I & II & III & IV & V & VI & Total \\
    \hline
    0.0--0.1 & 2.0 & 0.7 & 1.1 & 0.8 & 2.1 & 2.8 & 4.3 \\
    0.1--0.3 & 1.7 & 0.7 & 1.0 & 1.2 & 1.8 & 2.0 & 3.7 \\
    0.3--0.7 & 1.5 & 0.5 & 1.3 & 1.4 & 1.8 & 2.1 & 3.7 \\
    0.7--0.9 & 2.0 & 2.3 & 1.3 & 2.6 & 1.7 & 4.8 & 6.7 \\
    0.9--1.1 & 2.9 & 3.8 & 1.4 & 2.9 & 1.8 & 6.4 & 8.8 \\
    1.1--1.5 & 2.8 & 3.2 & 1.6 & 3.6 & 2.0 & 7.2 & 9.5 \\
    \hline \hline
  \end{tabular}
  \caption{Systematic uncertainties (expressed as percentages) on the ratio of charged-current inclusive \numu differential cross sections $\frac{d\sigma^{Fe}}{d\Xbj}/\frac{d\sigma^{CH}}{d\Xbj}$ with respect to \Xbj associated with  (I) subtraction of CH contamination, (II) detector response to muons and hadrons, (III) neutrino interaction models, (IV) final state interaction models, (V) flux and target number, and (VI) statistics.  The rightmost column shows the total uncertainty due to all sources.}
  \label{tab:x_ratio_26_92_sys_errors_example}
\end{table}
\endgroup
%%%
Muon and recoil energy reconstruction uncertainties are described in Ref.~\cite{nubarprl} and Ref.~\cite{nuprl}, respectively.
We evaluate systematic error from cross section and final state interaction models 
by varying underlying model parameters in GENIE 
within their uncertainties~\cite{Andreopoulos201087}. 
Since variations in model parameters affect calorimetric scale factors, these are reextracted during systematic error evaluation.  
Recoil energy and final state interaction model uncertainties increase with \Xbj, because interactions of lower energy hadrons are not as well constrained.  
An assay of detector components yields an uncertainty in scintillator, carbon, iron, and lead masses of 1.4\%, 0.5\%, 1.0\%, and 0.5\%, respectively.
The resulting uncertainties on  $\frac{d\sigma^{Fe}}{d\Xbj}/\frac{d\sigma^{CH}}{d\Xbj}$ are shown in Table~\ref{tab:x_ratio_26_92_sys_errors_example}\SuppNote{See Supplemental Material\SuppLocation for uncertainties on all cross section ratios as functions of \Enu and \Xbj}.

%% file: results-NukeCC.tex
%%\section{Results}

%%%%%%%%%%%%%%%%%%%%%
\begin{figure}[!htpb]
\centering
\includegraphics[trim = 5mm 12mm 18mm 8mm, clip, width=0.48\columnwidth]{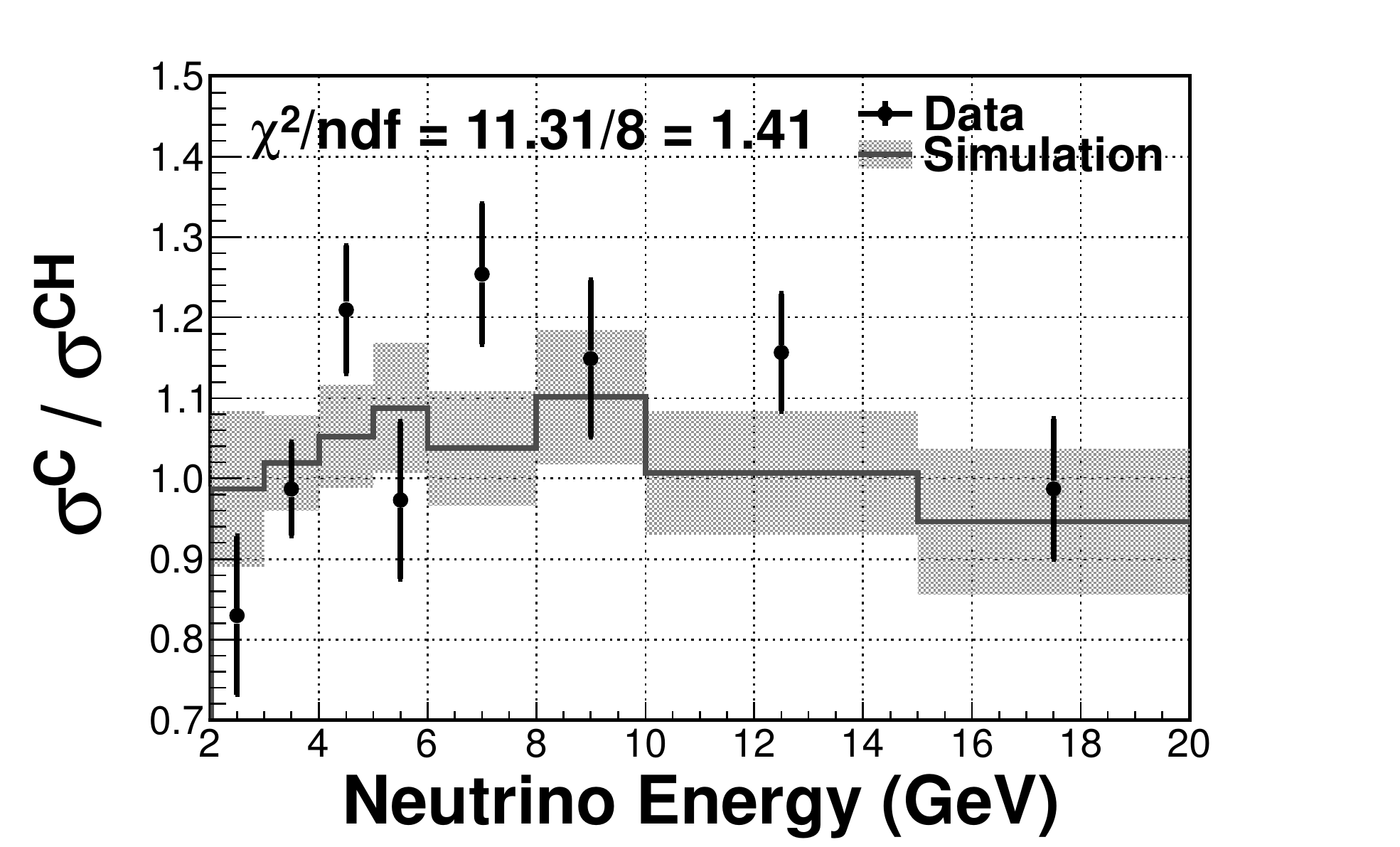}
\includegraphics[trim = 1mm 12mm 18mm 8mm, clip, width=0.48\columnwidth]{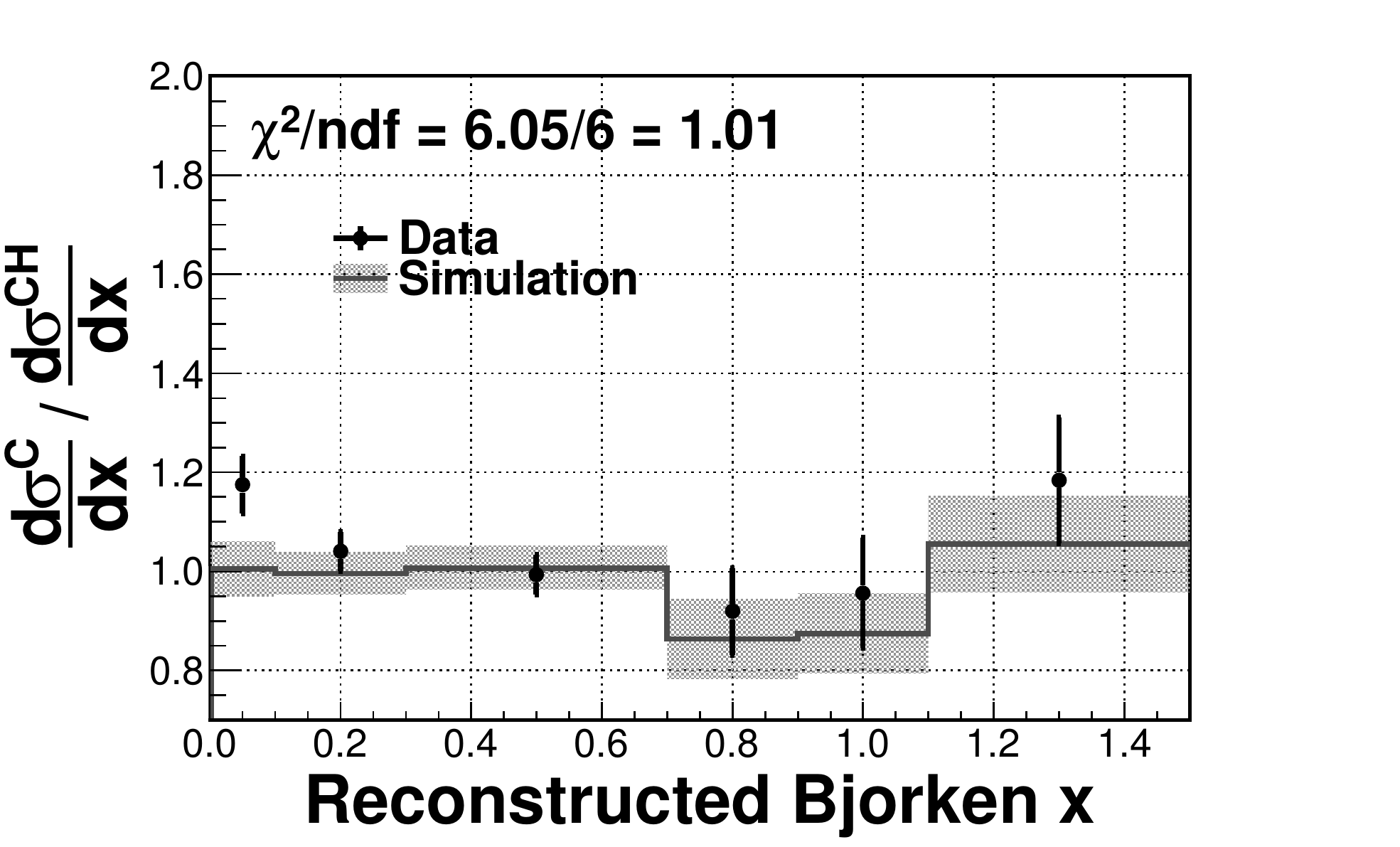}
\includegraphics[trim = 5mm 12mm 18mm 8mm, clip, width=0.48\columnwidth]{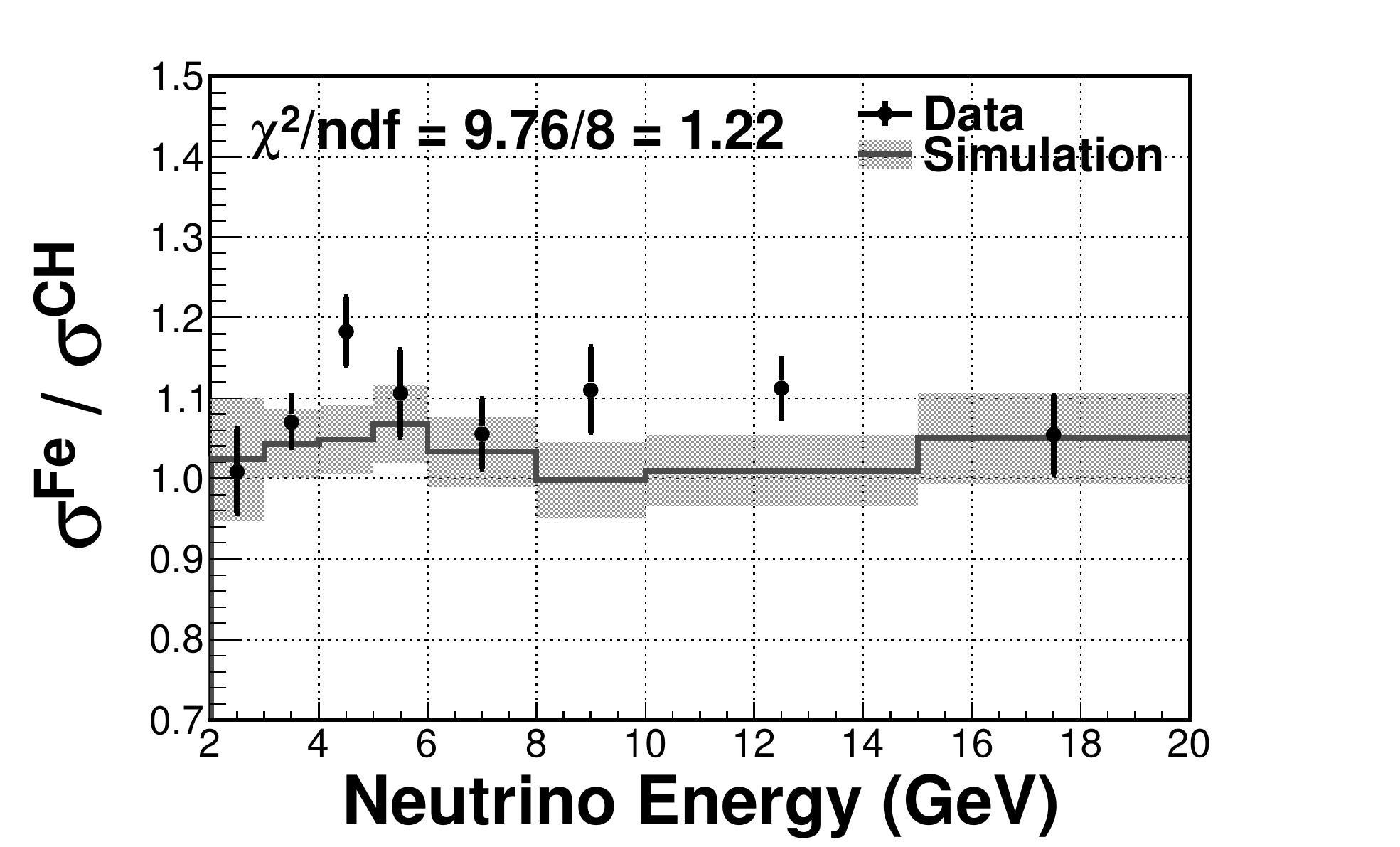}
\includegraphics[trim = 1mm 12mm 18mm 8mm, clip, width=0.48\columnwidth]{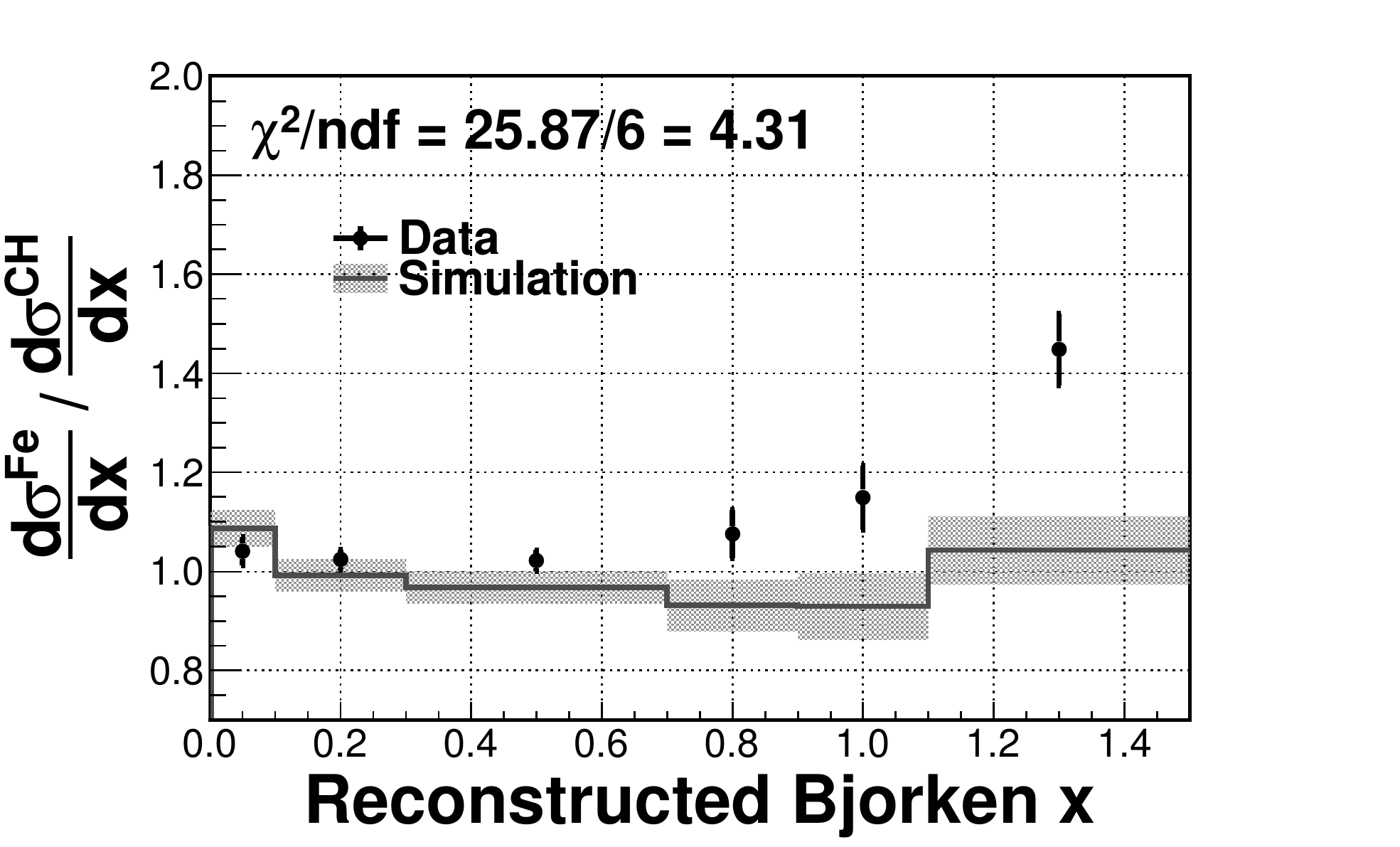}
\includegraphics[trim = 5mm 00mm 18mm 8mm, clip, width=0.48\columnwidth]{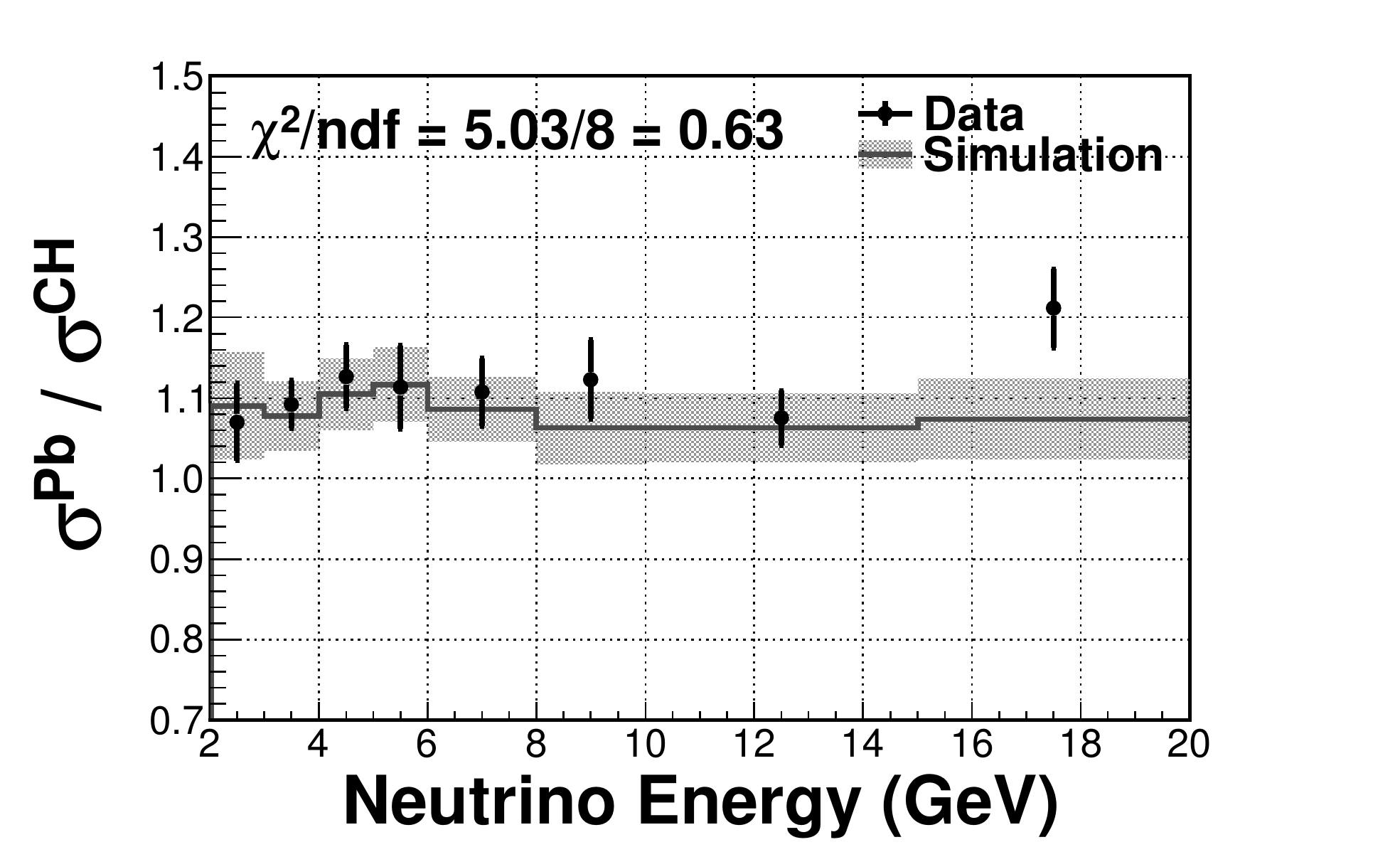}
\includegraphics[trim = 1mm 00mm 18mm 8mm, clip, width=0.48\columnwidth]{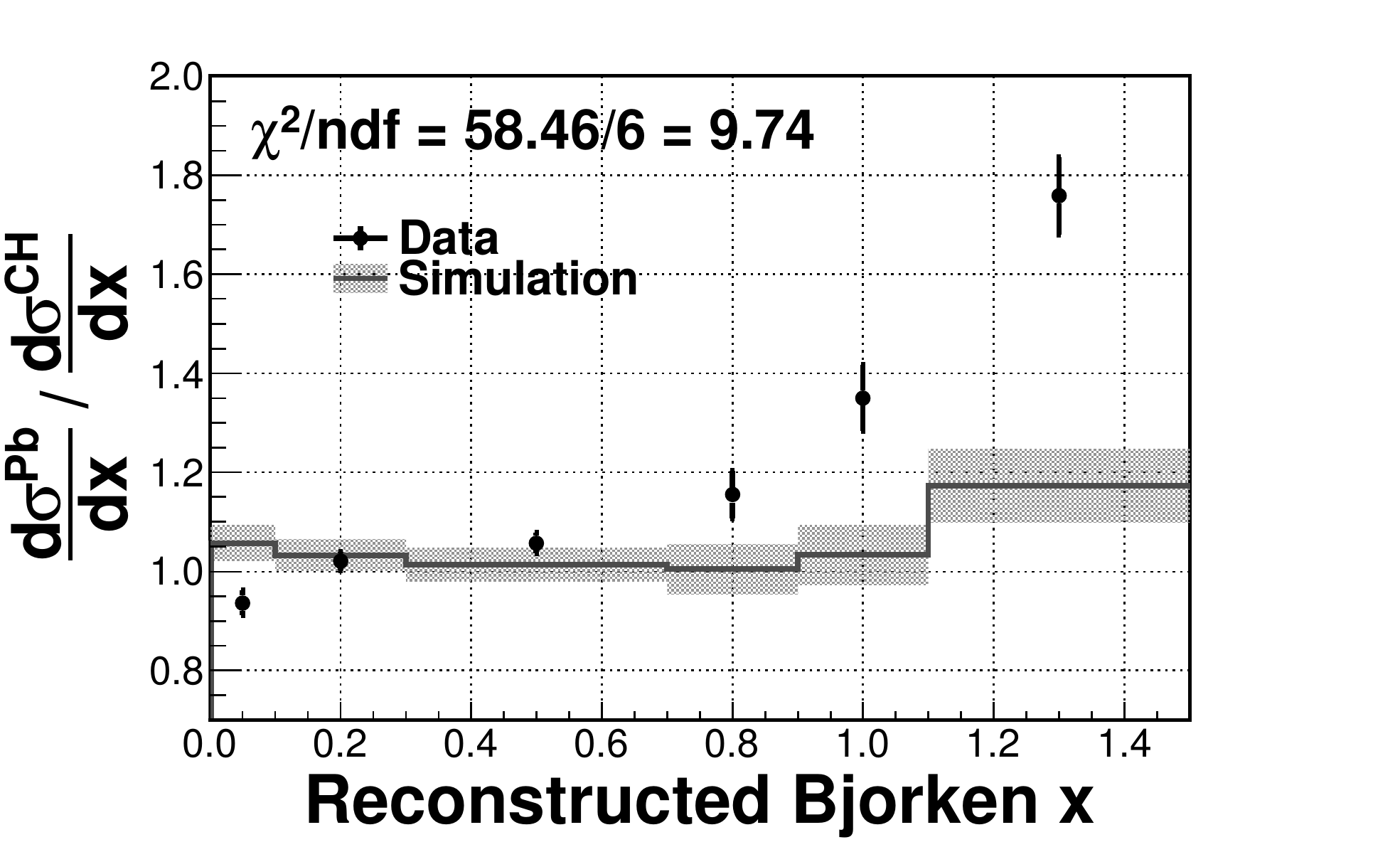}
\caption{Ratios of the charged-current inclusive \numu cross section per nucleon as a function of \Enu (left) and as a function of reconstructed \Xbj (right)
for C/CH (top), Fe/CH (middle), and Pb/CH (bottom).  Error bars on the data (simulation) show the statistical (systematic) uncertainties.  The $\chi^{2}$ calculation includes correlations among all bins shown.  Events with \Xbj greater than 1.5 are not shown.}
\label{fig:xsec_ratio_plots}
\end{figure}
%%%%%%%%%%%%%%%%%%%%%
Ratios of charged-current \numu cross sections per nucleon $\sigma\left(\Enu\right)$ and $\frac{d\sigma}{d\Xbj}$ are shown in Fig.~\ref{fig:xsec_ratio_plots}\SuppNote{See Supplemental Material\SuppLocation for cross section ratio measurements compared to simulation in tabular form and correlations of uncertainties among bins}.
Simulation reproduces these measurements within roughly 10\%.  
In contrast, measurements of $\frac{d\sigma^{\A}}{d\Xbj}/\frac{d\sigma^{CH}}{d\Xbj}$ show a suppression of the ratio compared to simulation at low \Xbj and an enhancement at high \Xbj, both of which increase with the size of the nucleus.

Low \Xbj bins are expected to show shadowing, which lowers the cross section for heavier nuclei~\cite{e665,Allport1989417,Kopeliovich:2012kw}.
Shadowing in these data may be larger than predicted for several reasons.
First, our data are at low \Qsq in the nonperturbative range (80\% of events below \unit[1.0]{\GeVcsq} and 60\% below \unit[0.5]{\GeVcsq}), while the model is tuned to data at much higher \Qsq where shadowing is well measured.
Second, shadowing in the model is assumed to be the same for C and Pb and equal to measurements from Fe~\cite{Bodek:2004pc}.
Finally, the shadowing model used for comparison is based on charged lepton data, which do not have axial-vector contributions.
The array of nuclear models available to modern neutrino experiments give 
similar results for these cross section ratios; none of which is confirmed by the data.

Higher \Xbj bins contain mostly ($\textgreater63\%$) quasielastic events, 
whose rates may be enhanced by meson-exchange currents~\cite{Carlson:2001mp,PhysRevC.80.065501,PhysRevC.81.045502,Bodek:2011ps,Amaro2011151,PhysRevC.83.045501,PhysRevD.88.113007},
which are not in the simulation.
The excess observed here may be related to the excess in MINOS Fe data at low inelasticity compared to a simulation with nuclear corrections based on lighter nuclei similar to GENIE's~\cite{PhysRevD.77.072002,arxiv:0708.1495}.
The failure of nuclear scaling models in this region has profound implications for neutrino oscillation experiments that utilize quasielastic events.
For example, T2K~\cite{PhysRevLett.111.211803,PhysRevLett.112.061802} must apply a nuclear model to relate the rate in the carbon of a near detector to oxygen in the far detector.  
LBNE~\cite{lbne_new} must extrapolate existing data on C, Fe, Pb to Ar.
Until better models exist that cover the relevant kinematic domain, oscillation experiments must 
incorporate the discrepancies measured here in evaluating systematic uncertainties.
More theoretical work is needed to correctly model nuclear effects in neutrino interactions, from the quasielastic to the deep inelastic regime.

%% file: acknowledgments.tex
\begin{acknowledgments}

This work was supported by the Fermi National Accelerator Laboratory
under U.S. Department of Energy Contract
No.\@ DE-AC02-07CH11359 which included the \minerva construction project.
Construction support also
was granted by the United States National Science Foundation under
Grant No. PHY-0619727 and by the University of Rochester. Support for
participating scientists was provided by NSF and DOE (USA) by CAPES
and CNPq (Brazil), by CoNaCyT (Mexico), by CONICYT (Chile), by
CONCYTEC, DGI-PUCP and IDI/IGI-UNI (Peru), by Latin American Center for
Physics (CLAF), by the Swiss National Science Foundation, and by RAS and the Russian Ministry of Education and Science (Russia).  We
thank the MINOS Collaboration for use of its
near detector data. Finally, we thank the staff of
Fermilab for support of the beam line and detector.

\end{acknowledgments}

%% file: biblio-NukeCC-bibtex.tex
\bibliographystyle{apsrev4-1}
\bibliography{CCInclNukeTarget}

%% file: appendix-NukeCC.tex
\section{Supplemental Material}
\input{full-dataMC-tables.tex}

\input{target-number-table.tex}
\input{nu-flux-table.tex}

\input{full-xsec-corr-tables.tex}

\input{full-xsec-sys-error-tables.tex}

\input{migration-tables.tex}
\input{model-compare-table.tex}

%% file: full-dataMC-tables.tex
%%%%%%%%%%%%%%%%%%%%%%%%%%%%%%%%%%%%%%%%%%%%%%%%%%%%
% These tables are made with RatioResultsTableDataMC
% http://cdcvs.fnal.gov/cgi-bin/public-cvs/cvsweb-public.cgi/AnalysisFramework/Ana/NukeCCInclusive/ana/make_tables/RatioResultTableDataMC.cxx?cvsroot=mnvsoft
%
  % 1. Change the value of var near the top of the code to "Enu"
  % 2. Compile (make)
% 3. Run (./RatioResultTableDataMC)
  % 4. Copy output to here
  % 5. Repeat steps 1--5 but with var="x"
  %%%%%%%%%%%%%%%%%%%%%%%%%%%%%%%%%%%%%%%%%%%%%%%%%%%%

\begin{table}[!ht]
  \begin{center}
    \begin{tabular}{ccccccc}
      \hline\hline
      & \Xbj Bin & Data & Sim. & $\sigma_{stat}$ & $\sigma_{sys}$ & $\frac{(\text{Data}-\text{Sim.})}{\sigma}$ \\ 

      \hline
      Carbon & 0.0--0.1 & 1.17 & 1.01 & 0.056 & 0.056 & 2.01 \\ 
       & 0.1--0.3 & 1.04 & 1.00 & 0.038 & 0.039 & 0.76 \\ 
       & 0.3--0.7 & 0.99 & 1.01 & 0.039 & 0.038 & -0.25 \\ 
       & 0.7--0.9 & 0.92 & 0.86 & 0.087 & 0.072 & 0.46 \\ 
       & 0.9--1.1 & 0.96 & 0.88 & 0.111 & 0.066 & 0.58 \\ 
       & 1.1--1.5 & 1.18 & 1.06 & 0.126 & 0.089 & 0.79 \\ 

      \hline
      Iron & 0.0--0.1 & 1.04 & 1.09 & 0.028 & 0.032 & -1.02 \\ 
       & 0.1--0.3 & 1.02 & 0.99 & 0.020 & 0.031 & 0.83 \\ 
       & 0.3--0.7 & 1.02 & 0.97 & 0.021 & 0.032 & 1.37 \\ 
       & 0.7--0.9 & 1.08 & 0.93 & 0.048 & 0.053 & 1.91 \\ 
       & 0.9--1.1 & 1.15 & 0.93 & 0.064 & 0.075 & 2.14 \\ 
       & 1.1--1.5 & 1.45 & 1.04 & 0.072 & 0.086 & 3.51 \\ 

      \hline
      Lead & 0.0--0.1 & 0.94 & 1.06 & 0.025 & 0.029 & -2.99 \\ 
       & 0.1--0.3 & 1.02 & 1.03 & 0.018 & 0.030 & -0.32 \\ 
       & 0.3--0.7 & 1.06 & 1.01 & 0.020 & 0.034 & 1.09 \\ 
       & 0.7--0.9 & 1.15 & 1.00 & 0.047 & 0.051 & 2.06 \\ 
       & 0.9--1.1 & 1.35 & 1.03 & 0.067 & 0.070 & 3.16 \\ 
       & 1.1--1.5 & 1.76 & 1.17 & 0.077 & 0.103 & 4.45 \\ 

      \hline\hline
    \end{tabular}
  \end{center}
  \caption{Comparison of measured values to simulation predictions for $\frac{d\sigma^{A}}{d\Xbj}/\frac{d\sigma^{CH}}{d\Xbj}$ for each \Xbj bin.}
  \label{tab:x_results_table}
\end{table}

\begin{table}[!ht]
\begin{center}
\begin{tabular}{ccccccc}
\hline\hline
& \Enu Bin (\GeV) & Data & Sim. & $\sigma_{stat}$ & $\sigma_{sys}$ & $\frac{(\text{Data}-\text{Sim.})}{\sigma}$ \\ 

\hline
Carbon & 2--3 & 0.83 & 0.99 & 0.096 & 0.072 & -1.23 \\ 
 & 3--4 & 0.99 & 1.02 & 0.056 & 0.051 & -0.41 \\ 
 & 4--5 & 1.21 & 1.05 & 0.077 & 0.063 & 1.51 \\ 
 & 5--6 & 0.97 & 1.09 & 0.096 & 0.058 & -0.95 \\ 
 & 6--8 & 1.25 & 1.04 & 0.084 & 0.070 & 1.85 \\ 
 & 8--10 & 1.15 & 1.10 & 0.093 & 0.068 & 0.38 \\ 
 & 10--15 & 1.16 & 1.01 & 0.069 & 0.075 & 1.36 \\ 
 & 15--20 & 0.99 & 0.95 & 0.084 & 0.076 & 0.33 \\ 

\hline
Iron & 2--3 & 1.01 & 1.02 & 0.051 & 0.072 & -0.17 \\ 
 & 3--4 & 1.07 & 1.04 & 0.031 & 0.041 & 0.50 \\ 
 & 4--5 & 1.18 & 1.05 & 0.041 & 0.043 & 2.17 \\ 
 & 5--6 & 1.11 & 1.07 & 0.052 & 0.042 & 0.54 \\ 
 & 6--8 & 1.06 & 1.03 & 0.042 & 0.038 & 0.37 \\ 
 & 8--10 & 1.11 & 1.00 & 0.051 & 0.043 & 1.57 \\ 
 & 10--15 & 1.11 & 1.01 & 0.035 & 0.043 & 1.73 \\ 
 & 15--20 & 1.05 & 1.05 & 0.046 & 0.049 & 0.06 \\ 

\hline
Lead & 2--3 & 1.07 & 1.09 & 0.046 & 0.062 & -0.25 \\ 
 & 3--4 & 1.09 & 1.08 & 0.029 & 0.042 & 0.27 \\ 
 & 4--5 & 1.13 & 1.10 & 0.038 & 0.042 & 0.37 \\ 
 & 5--6 & 1.11 & 1.12 & 0.050 & 0.039 & -0.05 \\ 
 & 6--8 & 1.11 & 1.09 & 0.040 & 0.035 & 0.38 \\ 
 & 8--10 & 1.12 & 1.06 & 0.047 & 0.039 & 0.91 \\ 
 & 10--15 & 1.08 & 1.06 & 0.032 & 0.038 & 0.23 \\ 
 & 15--20 & 1.21 & 1.07 & 0.046 & 0.048 & 1.92 \\ 

\hline\hline
\end{tabular}
\end{center}
\caption{Comparison of measured values to simulation predictions for $\sigma^{A}/\sigma^{CH}$ for each \Enu bin.}
\label{tab:enu_results_table}
\end{table}

%% file: target-number-table.tex
%%%%%%%%%%%%%%%%%
% Target number table
%%%%%%%%%%%%%%%%%
\begingroup
\squeezetable
\begin{table}
\begin{tabular}{cccccc}
\hline \hline
& Mass  & Protons             & Neutrons           & Nucleons & Uncertainty\\
Target & (ton) &  ($\times 10^{30}$) & ($\times 10^{30}$) & ($\times 10^{30}$) & \% \\
    \hline
    C  & 0.159 & 0.048 & 0.048 & 0.096 & 1.4 \\
    Fe & 0.628 & 0.176 & 0.203 & 0.379 & 0.5 \\
    Pb & 0.711 & 0.169 & 0.258 & 0.427 & 1.0 \\
    CH & 5.476 & 1.760 & 1.534 & 3.294 & 0.5 \\
      \hline \hline
      \end{tabular}
      \caption{Mass and uncertainty on mass; and number of protons, neutrons, and the total target nucleons in the fiducial volume for each nuclear target.}
      \label{tab:NumberOfTargets}
      \end{table}
      \endgroup

%% file: nu-flux-table.tex
\begingroup
\squeezetable
\begin{table}
  \begin{tabular}{cccccccccc}
    \hline \hline
    \Enu in Bin (\GeV) & 2--2.5 & 2.5--3 & 3--3.5 & 3.5--4 & 4--4.5 & 4.5--5 & 5--5.5 & 5.5--6 & 6--6.5 \\
    \numu Flux (neutrinos/\cmsq/POT)$\times10^{-8}$ & 0.409 & 0.501 & 0.526 & 0.419 & 0.253 & 0.137 & 0.080 & 0.055 & 0.042 \\
    \hline
    \Enu in Bin (\GeV) & 6.5--7 & 7--7.5 & 7.5--8 & 8--8.5 & 8.5--9 & 9--9.5 & 9.5--10 & 10--11 & 11--12 \\
    \numu Flux (neutrinos/\cmsq/POT)$\times10^{-8}$ & 0.036 & 0.030 & 0.027 & 0.024 & 0.021 & 0.019 & 0.017 & 0.030 & 0.025\\
    \hline
    \Enu in Bin (\GeV) & 12--13 & 13--14 & 14--15 & 15--16 & 16--17 & 17--18 & 18--19 & 19--20 & \\
    \numu Flux (neutrinos/\cmsq/POT)$\times10^{-8}$ & 0.021 & 0.018 & 0.015 & 0.012 & 0.010 & 0.009 & 0.007 & 0.006 &  \\
    \hline \hline
  \end{tabular}
  \caption{The calculated muon neutrino flux per proton on target (POT) for the data included in the analysis.}
  \label{tab:nu_flux_fine}
  \end{table}
\endgroup

%% file: full-xsec-corr-tables.tex
\begingroup
\squeezetable
\begin{table}
  \begin{tabular}{c|cccccc}
    \hline \hline
    \Xbj bin & 0.0--0.1 & 0.1--0.3 & 0.3--0.7 & 0.7--0.9 & 0.9--1.1 & 1.1--1.5 \\ 
    \hline
    Ratio of cross sections & 1.175 & 1.040 & 0.993 & 0.919 & 0.956 & 1.184 \\ 
    Error on ratio & $ \pm 0.074 $ & $ \pm 0.054 $ & $ \pm 0.055 $ & $ \pm 0.110 $ & $ \pm 0.126 $ & $ \pm 0.149 $ \\ 
    \hline
    \Xbj bin &  &  &  &  &  &  \\ 
    0.0--0.1 & 1.000 & 0.329 & 0.264 & 0.099 & 0.140 & 0.128 \\ 
    0.1--0.3 &  & 1.000 & 0.338 & 0.152 & 0.162 & 0.148 \\ 
    0.3--0.7 &  &  & 1.000 & 0.172 & 0.165 & 0.172 \\ 
    0.7--0.9 &  &  &  & 1.000 & 0.046 & -0.020 \\ 
    0.9--1.1 &  &  &  &  & 1.000 & 0.123 \\ 
    1.1--1.5 &  &  &  &  &  & 1.000 \\ 
    \hline \hline
  \end{tabular}
  \caption{Measured ratio of charged-current inclusive \numu differential cross sections $\frac{d\sigma^{C}}{d\Xbj}/\frac{d\sigma^{CH}}{d\Xbj}$ with respect to \Xbj, their total (statistical and systematic) uncertainties, and the correlation matrix for these uncertainties.}
  \label{tab:x_ratio_06_91_results_table}
\end{table}
\endgroup

\begingroup
\squeezetable
\begin{table}
  \begin{tabular}{c|cccccc}
    \hline \hline
    \Xbj bin & 0.0--0.1 & 0.1--0.3 & 0.3--0.7 & 0.7--0.9 & 0.9--1.1 & 1.1--1.5 \\ 
    \hline
    Ratio of cross sections & 1.041 & 1.024 & 1.022 & 1.076 & 1.150 & 1.449 \\ 
    Error on ratio & $ \pm 0.043 $ & $ \pm 0.037 $ & $ \pm 0.037 $ & $ \pm 0.067 $ & $ \pm 0.088 $ & $ \pm 0.095 $ \\ 
    \hline
    \Xbj bin &  &  &  &  &  &  \\ 
    0.0--0.1 & 1.000 & 0.525 & 0.457 & 0.247 & 0.262 & 0.256 \\ 
    0.1--0.3 &  & 1.000 & 0.534 & 0.243 & 0.341 & 0.290 \\ 
    0.3--0.7 &  &  & 1.000 & 0.393 & 0.377 & 0.372 \\ 
    0.7--0.9 &  &  &  & 1.000 & 0.128 & 0.204 \\ 
    0.9--1.1 &  &  &  &  & 1.000 & 0.359 \\ 
    1.1--1.5 &  &  &  &  &  & 1.000 \\ 
    \hline \hline
  \end{tabular}
  \caption{Measured ratio of charged-current inclusive \numu differential cross sections $\frac{d\sigma^{Fe}}{d\Xbj}/\frac{d\sigma^{CH}}{d\Xbj}$ with respect to \Xbj, their total (statistical and systematic) uncertainties, and the correlation matrix for these uncertainties.}
  \label{tab:x_ratio_26_92_results_table}
\end{table}
\endgroup

\begingroup
\squeezetable
\begin{table}
  \begin{tabular}{c|cccccc}
    \hline \hline
    \Xbj bin & 0.0--0.1 & 0.1--0.3 & 0.3--0.7 & 0.7--0.9 & 0.9--1.1 & 1.1--1.5 \\ 
    \hline
    Ratio of cross sections & 0.936 & 1.021 & 1.057 & 1.155 & 1.350 & 1.758 \\ 
    Error on ratio & $ \pm 0.041 $ & $ \pm 0.035 $ & $ \pm 0.038 $ & $ \pm 0.065 $ & $ \pm 0.085 $ & $ \pm 0.103 $ \\ 
    \hline
    \Xbj bin &  &  &  &  &  &  \\ 
    0.0--0.1 & 1.000 & 0.592 & 0.486 & 0.332 & 0.271 & 0.257 \\ 
    0.1--0.3 &  & 1.000 & 0.608 & 0.415 & 0.345 & 0.309 \\ 
    0.3--0.7 &  &  & 1.000 & 0.445 & 0.393 & 0.389 \\ 
    0.7--0.9 &  &  &  & 1.000 & 0.262 & 0.289 \\ 
    0.9--1.1 &  &  &  &  & 1.000 & 0.325 \\ 
    1.1--1.5 &  &  &  &  &  & 1.000 \\ 
    \hline \hline
  \end{tabular}
  \caption{Measured ratio of charged-current inclusive \numu differential cross sections $\frac{d\sigma^{Pb}}{d\Xbj}/\frac{d\sigma^{CH}}{d\Xbj}$ with respect to \Xbj, their total (statistical and systematic) uncertainties, and the correlation matrix for these uncertainties.}
  \label{tab:x_ratio_82_93_results_table}
\end{table}
\endgroup

\begingroup
\squeezetable
\begin{table}
  \begin{tabular}{c|cccccccc}
    \hline \hline
    \Enu (\GeV) bin & 2--3 & 3--4 & 4--5 & 5--6 & 6--8 & 8--10 & 10--15 & 15--20 \\ 
    \hline
    Ratio of cross sections & 0.830 & 0.987 & 1.210 & 0.973 & 1.254 & 1.149 & 1.157 & 0.987 \\ 
    Error on ratio & $ \pm 0.128 $ & $ \pm 0.077 $ & $ \pm 0.095 $ & $ \pm 0.115 $ & $ \pm 0.102 $ & $ \pm 0.114 $ & $ \pm 0.095 $ & $ \pm 0.111 $ \\ 
    \hline
    \Enu (\GeV) bin &  &  &  &  &  &  &  &  \\ 
    2--3 & 1.000 & 0.272 & 0.180 & 0.051 & 0.107 & 0.052 & 0.117 & 0.046 \\ 
    3--4 &  & 1.000 & 0.281 & 0.193 & 0.181 & 0.156 & 0.197 & 0.169 \\ 
    4--5 &  &  & 1.000 & 0.149 & 0.191 & 0.112 & 0.202 & 0.174 \\ 
    5--6 &  &  &  & 1.000 & 0.090 & 0.156 & 0.177 & 0.153 \\ 
    6--8 &  &  &  &  & 1.000 & 0.104 & 0.211 & 0.191 \\ 
    8--10 &  &  &  &  &  & 1.000 & 0.193 & 0.179 \\ 
    10--15 &  &  &  &  &  &  & 1.000 & 0.280 \\ 
    15--20 &  &  &  &  &  &  &  & 1.000 \\ 
    \hline \hline
  \end{tabular}
  \caption{Measured ratio of charged-current inclusive \numu total cross sections $\sigma^{C}/\sigma^{CH}$ as a function of \Enu, their total (statistical and systematic) uncertainties, and the correlation matrix for these uncertainties.}
  \label{tab:Enu_ratio_06_91_results_table}
\end{table}
\endgroup

\begingroup
\squeezetable
\begin{table}
  \begin{tabular}{c|cccccccc}
    \hline \hline
    \Enu (\GeV) bin & 2--3 & 3--4 & 4--5 & 5--6 & 6--8 & 8--10 & 10--15 & 15--20 \\ 
    \hline
    Ratio of cross sections & 1.009 & 1.070 & 1.183 & 1.106 & 1.056 & 1.110 & 1.113 & 1.055 \\ 
    Error on ratio & $ \pm 0.089 $ & $ \pm 0.051 $ & $ \pm 0.056 $ & $ \pm 0.066 $ & $ \pm 0.056 $ & $ \pm 0.064 $ & $ \pm 0.053 $ & $ \pm 0.067 $ \\ 
    \hline
    \Enu (\GeV) bin &  &  &  &  &  &  &  &  \\ 
    2--3 & 1.000 & 0.381 & 0.219 & 0.239 & 0.095 & 0.122 & 0.105 & 0.003 \\ 
    3--4 &  & 1.000 & 0.450 & 0.352 & 0.284 & 0.200 & 0.207 & 0.120 \\ 
    4--5 &  &  & 1.000 & 0.291 & 0.282 & 0.217 & 0.224 & 0.166 \\ 
    5--6 &  &  &  & 1.000 & 0.221 & 0.190 & 0.235 & 0.161 \\ 
    6--8 &  &  &  &  & 1.000 & 0.224 & 0.299 & 0.298 \\ 
    8--10 &  &  &  &  &  & 1.000 & 0.298 & 0.228 \\ 
    10--15 &  &  &  &  &  &  & 1.000 & 0.340 \\ 
    15--20 &  &  &  &  &  &  &  & 1.000 \\ 
    \hline \hline
  \end{tabular}
  \caption{Measured ratio of charged-current inclusive \numu total cross sections $\sigma^{Fe}/\sigma^{CH}$ as a function of \Enu, their total (statistical and systematic) uncertainties, and the correlation matrix for these uncertainties.}
  \label{tab:Enu_ratio_26_92_results_table}
\end{table}
\endgroup

\begingroup
\squeezetable
\begin{table}
  \begin{tabular}{c|cccccccc}
    \hline \hline
    \Enu (\GeV) bin & 2--3 & 3--4 & 4--5 & 5--6 & 6--8 & 8--10 & 10--15 & 15--20 \\ 
    \hline
    Ratio of cross sections & 1.070 & 1.092 & 1.127 & 1.113 & 1.107 & 1.123 & 1.075 & 1.212 \\ 
    Error on ratio & $ \pm 0.078 $ & $ \pm 0.050 $ & $ \pm 0.056 $ & $ \pm 0.063 $ & $ \pm 0.053 $ & $ \pm 0.060 $ & $ \pm 0.050 $ & $ \pm 0.062 $ \\ 
    \hline
    \Enu (\GeV) bin &  &  &  &  &  &  &  &  \\ 
    2--3 & 1.000 & 0.465 & 0.275 & 0.262 & 0.168 & 0.139 & 0.080 & 0.091 \\ 
    3--4 &  & 1.000 & 0.506 & 0.376 & 0.330 & 0.203 & 0.197 & 0.134 \\ 
    4--5 &  &  & 1.000 & 0.348 & 0.338 & 0.234 & 0.259 & 0.194 \\ 
    5--6 &  &  &  & 1.000 & 0.265 & 0.214 & 0.216 & 0.194 \\ 
    6--8 &  &  &  &  & 1.000 & 0.252 & 0.317 & 0.249 \\ 
    8--10 &  &  &  &  &  & 1.000 & 0.329 & 0.273 \\ 
    10--15 &  &  &  &  &  &  & 1.000 & 0.365 \\ 
    15--20 &  &  &  &  &  &  &  & 1.000 \\ 
    \hline \hline
  \end{tabular}
  \caption{Measured ratio of charged-current inclusive \numu total cross sections $\sigma^{Pb}/\sigma^{CH}$ as a function of \Enu, their total (statistical and systematic) uncertainties, and the correlation matrix for these uncertainties.}
  \label{tab:Enu_ratio_82_93_results_table}
\end{table}
\endgroup

%% file: full-xsec-sys-error-tables.tex
%%%%%
% These tables are made with RatioResultErrorGroupTable
%http://cdcvs.fnal.gov/cgi-bin/public-cvs/cvsweb-public.cgi/AnalysisFramework/Ana/NukeCCInclusive/ana/make_tables/RatioResultErrorGroupTable.cxx?cvsroot=mnvsoft
% 1. edit the file at the "vars" variable to push_back only x.
% 2. make
% 3. ./RatioResultErrorGroupTable
% 4. pipe or copy output to file
% 5. repeat steps 1-4 for Enu
% 6. Be sure to expand the description of error meaning in the caption.
%%%%%

\clearpage

\begingroup
\squeezetable
\begin{table}
  \begin{tabular}{cccccccc}
    \hline \hline
    \Xbj & I & II & III & IV & V & VI & Total \\ 
    \hline
    0.0--0.1 & 3.5 & 1.1 & 0.9 & 2.1 & 2.1 & 5.6 & 7.4 \\ 
    0.1--0.3 & 2.8 & 0.6 & 0.9 & 1.4 & 1.8 & 3.8 & 5.4 \\ 
    0.3--0.7 & 2.3 & 0.8 & 1.6 & 1.8 & 1.7 & 3.9 & 5.5 \\ 
    0.7--0.9 & 3.0 & 4.8 & 1.6 & 3.0 & 1.6 & 8.7 & 11.0 \\ 
    0.9--1.1 & 3.4 & 2.5 & 1.6 & 3.5 & 2.0 & 11.1 & 12.6 \\ 
    1.1--1.5 & 3.2 & 5.7 & 2.2 & 3.0 & 2.5 & 12.6 & 14.9 \\ 
    \hline \hline
  \end{tabular}
  \caption{Systematic uncertainties (expressed as percentages) on the ratio of charged-current inclusive \numu differential cross sections $\frac{d\sigma^{C}}{d\Xbj}/\frac{d\sigma^{CH}}{d\Xbj}$ with respect to \Xbj associated with  (I) subtraction of CH contamination, (II) detector response to muons and hadrons, (III) neutrino interactions, (IV) final state interactions, (V) flux and target number, and (VI) statistics.  The rightmost column shows the total uncertainty due to all sources.}
  \label{tab:x_ratio_06_91_sys_errors}
\end{table}
\endgroup

\begingroup
\squeezetable
\begin{table}
  \begin{tabular}{cccccccc}
    \hline \hline
    \Xbj & I & II & III & IV & V & VI & Total \\ 
    \hline
    0.0--0.1 & 2.0 & 0.7 & 1.1 & 0.8 & 2.1 & 2.8 & 4.3 \\ 
    0.1--0.3 & 1.7 & 0.7 & 1.0 & 1.2 & 1.8 & 2.0 & 3.7 \\ 
    0.3--0.7 & 1.5 & 0.5 & 1.3 & 1.4 & 1.8 & 2.1 & 3.7 \\ 
    0.7--0.9 & 2.0 & 2.3 & 1.3 & 2.6 & 1.7 & 4.8 & 6.7 \\ 
    0.9--1.1 & 2.9 & 3.8 & 1.4 & 2.9 & 1.8 & 6.4 & 8.8 \\ 
    1.1--1.5 & 2.8 & 3.2 & 1.6 & 3.6 & 2.0 & 7.2 & 9.5 \\ 
    \hline \hline
  \end{tabular}
  \caption{Systematic uncertainties (expressed as percentages) on the ratio of charged-current inclusive \numu differential cross sections $\frac{d\sigma^{Fe}}{d\Xbj}/\frac{d\sigma^{CH}}{d\Xbj}$ with respect to \Xbj associated with  (I) subtraction of CH contamination, (II) detector response to muons and hadrons, (III) neutrino interactions, (IV) final state interactions, (V) flux and target number, and (VI) statistics.  The rightmost column shows the total uncertainty due to all sources.}
  \label{tab:x_ratio_26_92_sys_errors}
\end{table}
\endgroup

\begingroup
\squeezetable
\begin{table}
  \begin{tabular}{cccccccc}
    \hline \hline
    \Xbj & I & II & III & IV & V & VI & Total \\ 
    \hline
    0.0--0.1 & 2.2 & 0.7 & 1.0 & 1.1 & 1.8 & 2.5 & 4.1 \\ 
    0.1--0.3 & 1.9 & 0.7 & 1.1 & 1.1 & 1.6 & 1.8 & 3.5 \\ 
    0.3--0.7 & 1.6 & 0.7 & 1.5 & 1.6 & 1.6 & 2.0 & 3.8 \\ 
    0.7--0.9 & 2.5 & 1.5 & 1.5 & 2.5 & 1.7 & 4.7 & 6.5 \\ 
    0.9--1.1 & 2.6 & 2.5 & 1.6 & 2.8 & 2.1 & 6.7 & 8.5 \\ 
    1.1--1.5 & 3.0 & 3.5 & 1.9 & 4.2 & 1.9 & 7.7 & 10.3 \\ 
    \hline \hline
  \end{tabular}
  \caption{Systematic uncertainties (expressed as percentages) on the ratio of charged-current inclusive \numu differential cross sections $\frac{d\sigma^{Pb}}{d\Xbj}/\frac{d\sigma^{CH}}{d\Xbj}$ with respect to \Xbj associated with  (I) subtraction of CH contamination, (II) detector response to muons and hadrons, (III) neutrino interactions, (IV) final state interactions, (V) flux and target number, and (VI) statistics.  The rightmost column shows the total uncertainty due to all sources.}
  \label{tab:x_ratio_82_93_sys_errors}
\end{table}
\endgroup

\begingroup
\squeezetable
\begin{table}
  \begin{tabular}{cccccccc}
    \hline \hline
    \Enu (\GeV) & I & II & III & IV & V & VI & Total \\ 
    \hline
    2--3 & 3.4 & 5.2 & 4.2 & 2.6 & 3.1 & 9.6 & 12.8 \\ 
    3--4 & 3.1 & 1.0 & 2.7 & 2.2 & 2.2 & 5.6 & 7.7 \\ 
    4--5 & 3.4 & 1.6 & 2.3 & 2.3 & 2.3 & 7.7 & 9.5 \\ 
    5--6 & 4.0 & 3.0 & 2.1 & 1.9 & 2.9 & 9.6 & 11.5 \\ 
    6--8 & 3.9 & 2.4 & 1.9 & 1.5 & 2.6 & 8.4 & 10.2 \\ 
    8--10 & 4.6 & 2.3 & 2.1 & 1.4 & 3.1 & 9.3 & 11.4 \\ 
    10--15 & 5.2 & 1.2 & 1.9 & 2.0 & 2.6 & 6.9 & 9.5 \\ 
    15--20 & 5.9 & 1.6 & 1.9 & 1.7 & 2.9 & 8.4 & 11.1 \\ 
    \hline \hline
  \end{tabular}
  \caption{Systematic uncertainties (expressed as percentages) on the ratio of charged-current inclusive \numu total cross sections $\sigma^{C}/\sigma^{CH}$ as a function of \Enu associated with  (I) subtraction of CH contamination, (II) detector response to muons and hadrons, (III) neutrino interactions, (IV) final state interactions, (V) flux and target number, and (VI) statistics.  The rightmost column shows the total uncertainty due to all sources.}
  \label{tab:Enu_ratio_06_91_sys_errors}
\end{table}
\endgroup

\begingroup
\squeezetable
\begin{table}
  \begin{tabular}{cccccccc}
    \hline \hline
    \Enu (\GeV) & I & II & III & IV & V & VI & Total \\ 
    \hline
    2--3 & 1.7 & 5.1 & 3.9 & 1.8 & 2.3 & 5.1 & 8.9 \\ 
    3--4 & 1.5 & 0.5 & 2.5 & 1.8 & 2.0 & 3.1 & 5.1 \\ 
    4--5 & 1.7 & 0.9 & 1.9 & 1.7 & 2.1 & 4.1 & 5.6 \\ 
    5--6 & 2.0 & 1.2 & 1.8 & 1.5 & 2.3 & 5.2 & 6.6 \\ 
    6--8 & 2.0 & 1.4 & 1.5 & 1.1 & 2.2 & 4.2 & 5.6 \\ 
    8--10 & 2.2 & 0.9 & 1.7 & 1.1 & 2.3 & 5.1 & 6.4 \\ 
    10--15 & 2.2 & 0.6 & 2.0 & 1.0 & 2.2 & 3.5 & 5.3 \\ 
    15--20 & 2.9 & 1.5 & 2.2 & 1.3 & 2.5 & 4.6 & 6.7 \\ 
    \hline \hline
  \end{tabular}
  \caption{Systematic uncertainties (expressed as percentages) on the ratio of charged-current inclusive \numu total cross sections $\sigma^{Fe}/\sigma^{CH}$ as a function of \Enu associated with  (I) subtraction of CH contamination, (II) detector response to muons and hadrons, (III) neutrino interactions, (IV) final state interactions, (V) flux and target number, and (VI) statistics.  The rightmost column shows the total uncertainty due to all sources.}
  \label{tab:Enu_ratio_26_92_sys_errors}
\end{table}
\endgroup

\begingroup
\squeezetable
\begin{table}
  \begin{tabular}{cccccccc}
    \hline \hline
    \Enu (\GeV) & I & II & III & IV & V & VI & Total \\ 
    \hline
    2--3 & 1.5 & 3.8 & 3.7 & 2.0 & 2.1 & 4.6 & 7.8 \\ 
    3--4 & 1.4 & 0.5 & 2.8 & 1.9 & 1.8 & 2.9 & 5.0 \\ 
    4--5 & 2.0 & 1.0 & 2.2 & 1.8 & 1.9 & 3.8 & 5.6 \\ 
    5--6 & 1.9 & 1.0 & 1.9 & 1.6 & 2.1 & 5.0 & 6.3 \\ 
    6--8 & 1.8 & 0.7 & 1.6 & 1.2 & 2.0 & 4.0 & 5.3 \\ 
    8--10 & 2.0 & 0.6 & 1.7 & 1.3 & 2.1 & 4.7 & 6.0 \\ 
    10--15 & 2.3 & 0.9 & 1.8 & 0.8 & 2.0 & 3.2 & 5.0 \\ 
    15--20 & 2.6 & 0.8 & 2.1 & 0.8 & 2.2 & 4.6 & 6.2 \\ 
    \hline \hline
  \end{tabular}
  \caption{Systematic uncertainties (expressed as percentages) on the ratio of charged-current inclusive \numu total cross sections $\sigma^{Pb}/\sigma^{CH}$ as a function of \Enu associated with  (I) subtraction of CH contamination, (II) detector response to muons and hadrons, (III) neutrino interactions, (IV) final state interactions, (V) flux and target number, and (VI) statistics.  The rightmost column shows the total uncertainty due to all sources.}
  \label{tab:Enu_ratio_82_93_sys_errors}
\end{table}
\endgroup

%% file: migration-tables.tex
\begingroup
\squeezetable
\begin{table}
\begin{tabular}{c|ccccccc}
\hline \hline
\Xbj bin & 0.0--0.1 & 0.1--0.3 & 0.3--0.7 & 0.7--0.9 & 0.9--1.1 & 1.1--1.5 & overflow \\ 
\hline
0.0--0.1 & 73 & 23 & 3 & 0 & 0 & 0 & 0 \\ 
0.1--0.3 & 12 & 60 & 23 & 2 & 1 & 1 & 2 \\ 
0.3--0.7 & 4 & 20 & 47 & 9 & 5 & 6 & 9 \\ 
0.7--0.9 & 2 & 11 & 30 & 11 & 9 & 10 & 26 \\ 
0.9--1.1 & 2 & 8 & 30 & 12 & 6 & 10 & 31 \\ 
1.1--1.5 & 3 & 7 & 21 & 8 & 8 & 14 & 38 \\ 
\hline \hline
\end{tabular}
\caption{Fractional bin migration in variable \Xbj for the carbon sample as predicted by simulation.  The value of the bin is the percent of events that were generated in an \Xbj bin (row) that were reconstructed into an \Xbj bin (column).}
\label{tab:x_06_migration_table}
\end{table}
\endgroup

\begingroup
\squeezetable
\begin{table}
\begin{tabular}{c|ccccccc}
\hline \hline
\Xbj bin & 0.0--0.1 & 0.1--0.3 & 0.3--0.7 & 0.7--0.9 & 0.9--1.1 & 1.1--1.5 & overflow \\ 
\hline
0.0--0.1 & 73 & 23 & 3 & 0 & 0 & 0 & 0 \\ 
0.1--0.3 & 12 & 58 & 23 & 2 & 1 & 1 & 2 \\ 
0.3--0.7 & 3 & 18 & 47 & 10 & 6 & 6 & 9 \\ 
0.7--0.9 & 2 & 7 & 31 & 12 & 9 & 12 & 27 \\ 
0.9--1.1 & 2 & 6 & 23 & 12 & 9 & 15 & 34 \\ 
1.1--1.5 & 2 & 5 & 16 & 10 & 9 & 14 & 44 \\ 
\hline \hline
\end{tabular}
\caption{Fractional bin migration in variable \Xbj for the iron sample as predicted by simulation.  The value of the bin is the percent of events that were generated in an \Xbj bin (row) that were reconstructed into an \Xbj bin (column).}
\label{tab:x_26_migration_table}
\end{table}
\endgroup

\begingroup
\squeezetable
\begin{table}
\begin{tabular}{c|ccccccc}
\hline \hline
\Xbj bin & 0.0--0.1 & 0.1--0.3 & 0.3--0.7 & 0.7--0.9 & 0.9--1.1 & 1.1--1.5 & overflow \\ 
\hline
0.0--0.1 & 72 & 24 & 4 & 0 & 0 & 0 & 0 \\ 
0.1--0.3 & 12 & 59 & 23 & 2 & 1 & 1 & 1 \\ 
0.3--0.7 & 3 & 19 & 47 & 10 & 6 & 6 & 9 \\ 
0.7--0.9 & 2 & 8 & 29 & 13 & 10 & 12 & 25 \\ 
0.9--1.1 & 2 & 6 & 23 & 12 & 11 & 13 & 33 \\ 
1.1--1.5 & 2 & 5 & 16 & 11 & 8 & 14 & 44 \\ 
\hline \hline
\end{tabular}
\caption{Fractional bin migration in variable \Xbj for the lead sample as predicted by simulation.  The value of the bin is the percent of events that were generated in an \Xbj bin (row) that were reconstructed into an \Xbj bin (column).}
\label{tab:x_82_migration_table}
\end{table}
\endgroup

\begingroup
\squeezetable
\begin{table}
\begin{tabular}{c|ccccccc}
\hline \hline
\Xbj bin & 0.0--0.1 & 0.1--0.3 & 0.3--0.7 & 0.7--0.9 & 0.9--1.1 & 1.1--1.5 & overflow \\ 
\hline
0.0--0.1 & 75 & 22 & 2 & 0 & 0 & 0 & 0 \\ 
0.1--0.3 & 10 & 66 & 21 & 1 & 1 & 1 & 1 \\ 
0.3--0.7 & 2 & 15 & 59 & 11 & 5 & 4 & 3 \\ 
0.7--0.9 & 2 & 5 & 33 & 21 & 14 & 13 & 12 \\ 
0.9--1.1 & 2 & 4 & 19 & 19 & 17 & 19 & 20 \\ 
1.1--1.5 & 1 & 4 & 12 & 12 & 14 & 24 & 33 \\ 
\hline \hline
\end{tabular}
\caption{Fractional bin migration in variable \Xbj for the scintillator sample as predicted by simulation.  The value of the bin is the percent of events that were generated in an \Xbj bin (row) that were reconstructed into an \Xbj bin (column).}
\label{tab:x_99_migration_table}
\end{table}
\endgroup

%% file: model-compare-table.tex
\begingroup
\squeezetable
\begin{table}
\tiny
\begin{tabular}{c|cccc|cccc|cccc}
\hline \hline
& \multicolumn{4}{c}{C/CH} & \multicolumn{4}{c}{Fe/CH} & \multicolumn{4}{c}{Pb/CH} \\ 
\Xbj & G & $\sigma_{st}$ & KP & BY & G & $\sigma_{st}$ & KP & BY & G  & $\sigma_{st}$ & KP& BY\\
    &   & \%   & $\Delta$\%         & $\Delta$\%         &   & \%   & $\Delta$\%         &    $\Delta$\%     &    & \%   & $\Delta$\%         &  $\Delta$\%      \\
    \hline
    0.0--0.1 & 1.050 & 1.0 & 0.3 & 0.0 & 1.011 & 0.5 & -0.4 & 1.2 & 1.037 & 0.5 & -1.5 & 0.8 \\ 
    0.1--0.3 & 1.034 & 0.7 & -0.3 & 0.0 & 1.017 & 0.3 & -0.7 & -0.5 & 1.071 & 0.3 & -1.0 & -0.7 \\ 
    0.3--0.7 & 1.049 & 0.8 & -0.1 & 0.0 & 1.049 & 0.4 & 0.0 & 0.0 & 1.146 & 0.4 & 0.4 & 0.6 \\ 
    0.7--0.9 & 1.089 & 1.8 & -0.1 & 0.0 & 0.995 & 0.9 & 0.4 & 0.1 & 1.045 & 0.9 & 0.1 & 0.7 \\ 
    0.9--1.1 & 1.133 & 2.3 & -0.1 & 0.0 & 0.948 & 1.1 & 0.2 & 0.0 & 0.985 & 1.1 & 0.2 & 0.2 \\ 
    1.1--1.5 & 1.111 & 2.2 & 0.0 & 0.0 & 0.952 & 1.1 & 0.0 &  0.0 & 1.036 & 1.1 & 0.1 & 0.0 \\  
    \hline \hline 
    \end{tabular} 
    \caption{Predictions for charged-current cross section per nucleon ratios with $2<\Enu<20$~\GeV and $\thetamu<17^{\circ}$ from GENIE (G)~\cite{Bodek:2004pc} with associated statistical uncertainty.  Also shown is the deviation from GENIE predicted by the Kulagin-Petti (KP)~\cite{PhysRevD.76.094023,Kulagin2006126} and updated Bodek-Yang (BY)~\cite{by13} models for nuclear modification of nonresonant inelastic events.  Statistical uncertainty and deviations from GENIE are expressed as percentages.  The model deviations are calculated using event reweighting, thus there is no statistical variation among models.}
    \label{tab:models_table} 
    \end{table} 
    \endgroup